\documentclass[preprint,12pt]{elsarticle}
\usepackage{multirow}
\usepackage[]{amssymb}

\begin{document}

\begin{frontmatter}

\title{Effects of octahedral tilting on the electronic structure and optical
properties of $d^0$ double perovskites $\mathbf{\rm
A_2ScSbO_6}$ ($\mathbf{\rm A=Sr, Ca}$)}

      \author[label1,label2]{Rajyavardhan Ray}
      \address[label1]{IFW Dresden,
            Helmholtzstr. 20, D-01069, Dresden, Germany}
      \address[label2]{Dresden Center for Computational Material Science (DCMS), TU
            Dresden, 01062 Dresden, Germany}
      \ead{r.ray@ifw-dresden.de}

      \author[label3]{A K Himanshu}
      \address[label3]{Variable Energy Cyclotron Centre, 1/AF, Bidhannagar,
            Kolkata, India - 700064}
            \ead{akh@vecc.gov.in}
      \author[label3]{Pintu Sen}
      \author[label4]{Uday Kumar}
            \address[label4]{Department of Physical Sciences, Indian Institute of
            Science Education and Research Kolkata, Mohanpur,
            West Bengal, India - 741246}
      \author[label1,label2]{Manuel Richter}
      \author[label5]{T P Sinha}
            \address[label5]{Department of Physics, Bose Institute, 93/1, APC Road,
            Kolkata, India - 70009}

\begin{abstract}
 With increasing temperature, ${\rm Sr}_2{\rm ScSbO}_6$
 undergoes three structural phase transitions at approximately 
 ${\rm 400K}$, ${\rm 560K}$ and ${\rm 650K}$, leading to the following sequence 
 of phases: $P2_1/n \rightarrow I2/m \rightarrow I4/m \rightarrow
 Fm\bar{3}m$, making it an ideal candidate to study the
 effects of octahedral tilting keeping other parameters fixed. 
 To ascertain the isolated effects of octahedral distortions,
 the electronic and optical properties of the monoclinic $P2_1/n$
 (at room temperature), monoclinic $I2/m$ 
 (at ${\rm 430K}$), tetragonal $I4/m$ (at ${\rm
 613K}$) and the cubic $Fm\bar{3}m$ (at ${\rm 660K}$) phases 
 have been studied in terms of the electronic
 structure, dielectric constant, optical
 conductivity and electron energy loss spectroscopy using density
 functional theory. ${\rm Ca}_2{\rm ScSbO}_6$, on the other hand,
 shows only a $P2_1/n$ phase at room temperature and its properties have
 been been compared with the corresponding ${\rm Sr}$ compound.
 UV-Vis spectroscopic studies of the optical properties of the
 room-temperature phase of these $d^0$ double perovskite 
 have been performed and presence of large
 direct bandgap for both the compounds have been reported. 
 The electronic bandgaps for the room temperature phases is found to be in 
 good agreement with the corresponding experimental values obtained using the 
 Kubelka-Munk function.  Interestingly, in contrast to other Sc-based
 $d^0$ double perovskites, with increasing octahedral
 distortions, the effective $t_{\rm 2g}$ bandwidth remains
 unaffected while the states forming the band change due to changes
 in unit cell orientation, leading to small effects on
 the electronic and optical properties.
\end{abstract} 

\end{frontmatter}

\section{Introduction}
 Double perovskite systems, represented as ${\rm A_2BB'O_6}$, 
 have captivated current research interests due to the
 flexibility of their structure, allowing novel device applications
 originating from their low reactivity, good dielectric, magnetic and
 optical properties \cite{ReviewDP}.
 Of particular interest are the Antimony based double perovskite
 compounds which have attracted
 considerable attention of late \cite{CussenAntimonyInterest1,
 KashimaAntimonyInterest2, KarundasaAntimonyInterest3, PrimoMartinAntimonyInterest4,
 TauberAntimonyInterest5, FuAntimonyInterest6, LufasoAntimonyInterest7,
 IvanovAntimonyInterest8, BhartiTPSAntimonyInterest9,Faik2012}, due to possible technological
 applications \cite{TauberAntimonyInterest5}.
 Recently, two compounds ${\rm A_2ScSbO_6}$ (${\rm A=Sr, Ca}$) 
 have been synthesized and experimentally probed
 over a wide temperature range in order to study their crystal structure
 and phase transitions \cite{Faik2012}. 
 Both compounds are non-magnetic due to $+3$ and $+5$
 valency of the ${\rm Sc}$ and ${\rm Sb}$ cations, respectively, leading to
 unfilled Sc-$3d$ ($d^0$-ness) and fully-filled Sb-$4d$
 ($d$-full-ness) orbitals. 

 The electronic properties of some of the $d^0$-perovskites have been
 investigated earlier \cite{Eng2003}. In particular, changes in the electronic bandgap
 as function of the electronegativity of the transition metal ions and
 the conduction bandwidth have been reported. The focus of the study was on the effects
 arising out of changes in A-site cation and one of the B-site metal cations while keeping the
 $d^0$-metal cation fixed, for different $d^0$-metal cations. 
 It was shown that, with increasing
 octahedral tilting distortions, the conduction band narrows and the
 bandgap increases. It was also found that large electronegativity difference in
 the ${\rm B}$ and ${\rm B'}$-site cation leads to relativity larger
 effects on the electronic bandgaps. It was posited that a change of up to 
 2 eV in the electronic bandgap could be obtained by changing the chemical
 composition. While it is important to study the properties
 for different groups in the periodic table to ascertain the
 applicability of these materials, the role
 of octahedral distortions alone cannot be understood using this
 approach. ${\rm Sr}_2{\rm ScSbO}_6$ (SSS) is ideal to understand the
 isolated effects of octahedral tilting in $d^0$ double perovskites as it 
 undergoes multiple structural phase transitions. 

 Earlier, SSS has also been reported to be tetragonal with an unusually 
 high dielectric constant, $\epsilon = 8.8$ \cite{TauberAntimonyInterest5}.
 This implies a high refractive index $n \approx 3$ and that this
 material could be useful for a buffer compound as well as
 dielectric resonators and filters. 
 The recent synthesis, however, finds this compound to be monoclinic at
 room temperature and undergoing multiple structural phase transitions
 with increasing temperature \cite{Faik2012}.
 Also, with $d^0$-configuration of the Sc atom in ${\rm A_2ScSbO_6}$ (${\rm
 A=Sr, Ca})$, these compounds could be
 ferroelectric, exhibiting spontaneous
 polarization \cite{NatureReviewMultiferroics}.
 For potential technological applications in dielectric resonators and
 filters such as interference filters, optical fibres and reflective
 coating, accurate knowledge of 
 their optical response over a wide range of energy values and also as a
 function of the crystal geometry is required.
 Despite promising features, a detailed study of the electronic structure and
 optical properties of these compounds have not been
 carried out yet.
 Therefore, in this article, we
 focus on the electronic and optical properties of ${\rm Sr}_2{\rm
 ScSbO}_6$ (SSS) and ${\rm Ca}_2{\rm ScSbO}_6$ (CSS). 
 
 The primary aim of
 this article is two-fold: (i) to supplement the earlier studies on
 $d^0$-perovskites towards an improved understanding of the effects of octahedral tilting on the 
 electronic and optical properties, and (ii) to characterize and resolve, possibly
 rectify, the electronic, optical and ferroelectric properties of these compounds.

 With the general formula
 ${\rm A_2BB'O_6}$, double perovskites can be represented as a
 three-dimensional network of alternating octahedra ${\rm BO_6}$ and
 ${\rm B'O_6}$, with A-site atoms occupying the interstitial spaces between
 these octahedra. Typically, A-site atoms are alkaline earth metals such as
 Ba, Sr, Ca or a Lanthanide and the B-site atoms are transition metals.
 They constitute an important class of materials, characterized by
 structural distortions from the prototype cubic structure. These distortions 
 are caused by rotations of the ${\rm BO_6}$ and ${\rm B'O_6}$ octahedras to 
 accommodate the A-site cation.

 The ideal double perovskites have cubic
 symmetry with space group $Fm \bar3m$ (No. $225$). 
 A good measure of the deviation from this prototype cubic symmetry is
 the tolerance factor, defined as $t = (r_{\rm A} +
 r_O) / \sqrt{2} (\bar r_{B,B'} + r_O$), where $r_{A}$, $r_{O}$ and
 $r_{B,B'}$ are the ionic radii of A, O and the averaged
 ionic radius of the ${\rm B}$ and ${\rm B'}$ ions \cite{Megaw_tol}. Smallness of the A cation, which is
 the case for majority of double perovskites, leads to tilting of the
 octahedra. This further leads to stabilization of a crystal structure
 with lower symmetry \cite{Reany_tol}.
 For $t<1$, the structure is usually distorted from the prototype cubic
 symmetry.
 The tolerance factors of SSS and CSS is equal to 0.969 and 0.897
 respectively. This suggests
 that the room-temperature structures for both these compounds should not
 be cubic and that CSS should be more distorted than SSS, as also seen experimentally \cite{Faik2012}.

 With decreasing temperature, SSS undergoes three structural phase transformations:
 cubic $Fm\bar{3}m \rightarrow$ tetragonal $I4/m \rightarrow$ monoclinic
 $I2/m \rightarrow$ monoclinic $P2_1/n$, at approximately 650K, 560K and
 400K, respectively. The CSS compound, on the other hand, has a
 monoclinic structure (space group $P2_1/n$) at
 room temperature and shows no phase transformations up to temperature
 values as high as 1400K \cite{Faik2012}.

 The room-temperature monoclinic structure, space group $P2_1/n$, of
 these double perovskites
 is characterized by rock salt ordering of the ${\rm B}$ and
 ${\rm B'}$-site cations over alternating layers. The degree of cation ordering
 depends mostly on charge difference, size, and polarization of ${\rm
 B}$ and ${\rm B'}$-site cations. Due to the significantly large difference in the ionic
 radii of the ${\rm Sb^{5+}}$ and ${\rm Sc^{3+}}$ ions ($\Delta r = 0.145
 {\rm \AA}$), the ${\rm Sb}$ and ${\rm Sc}$ atoms are
 completely ordered in the two distinct B sites, in both SSS and CSS
 \cite{Faik2012}.

 As SSS undergoes structural phase transitions from
 cubic to monoclinic symmetry, we study the variation
 in the electronic bandgap, refractive index, and optical anisotropy across 
 these transitions using DFT to investigate the isolated effects of
 octahedral distortions. We also obtain the bandgaps of the room
 temperature phases of SSS and CSS using the diffuse reflectance
 spectroscopy and Kubelka-Munk function and compare the results with the
 corresponding theoretical results.
 This article is organized as
 follows: After mentioning the experimental and computational details in
 Sec. 2, we
 present and discuss our results in Sec. 3, and conclude in Sec. 4.

\section{Experimental and Computational Details}

\subsection{Synthesis and Diffuse Reflectance Measurements}
 Polycrystalline samples of ${\rm A_2ScSbO_6}$ (${\rm A=Sr, Ca}$) were
 prepared by solid state reaction method at room temperature, as described in Ref.
 \cite{Faik2012}. The bandgaps of the room temperature phases of these
 compounds were determined from the diffuse reflectance measurements in
 the UV-Vis range as suggested by Davis and Mott \cite{DavisMott}.
 The diffuse reflectance spectrum were obtained 
 using a Perkin-Elmer 950 UV/Vis/NIR spectrophotometer. It was later
 converted to an equivalent Kubelka-Munk (KM) absorption spectrum:
 \begin{equation}
       F(R_{\infty}) = \frac{(1-R_\infty)^2}{2R_{\infty}} =
       \frac{\alpha}{s}\,\,,
       \label{eqn:km}
 \end{equation}
 where $R_\infty= R_{\rm sample}/R_{\rm Spectralon}$ is the reflectance of the sample relative to a
 reference sample (Spectralon), $\alpha$ is the absorption coefficient
 and $s$ is the scattering coefficient. In the vicinity of the
 absorption edge, the energy dependence of the absorption coefficient
 can be expressed as \cite{DavisMott,linear}: 
 \begin{equation}
       \alpha = A\,\frac{(h\nu - E_{\rm g})^n}{h\nu}\,\,,
       \label{eqn:abs_edge}
 \end{equation}
 where $A$ is the proportionality constant, $h\nu$ is the energy of the incident photon and $E_{\rm g}$ is
 the optical bandgap energy. The values of $n$ can be appropriately
 chosen between $n=1/2$ and $2$ corresponding, respectively, to
 direct and indirect transitions. Thus, the bandgap can be obtained by
 inverting the above relation and, using Eq. (\ref{eqn:km}), plotting
 $[F(R_\infty) h\nu]^{1/n}$ with suitable $n$ as a function of energy. The intercept of the linear part of the spectrum near the
 absorption edge on the energy axis is equal to the bandgap.

\subsection{Computational Details}
 We employ the full-potential linear augmented plane wave (FPLAPW)
 method within the framework of density functional theory (DFT), as
 implemented in WIEN2k \cite{BlahaWIEN2k}, to calculate the electronic
 density of states (DOS), bandstructure, and optical response. 
 As the high temperature cubic $Fm \bar{3}m$ phase was not
 obtained by the authors of Ref. \cite{Faik2012} during the synthesis, we
 obtain the relaxed atomic positions by volume
 optimization in WIEN2k together with the force switch. 
 In this phase, we consider the estimated lattice parameter at 660 K
 \cite{Faik2012}. All the atomic positions, except for Oxygen atom, are
 fixed due to symmetry. The optimal atomic position
 for Oxygen was found to be ($0.2554,0,0$).

 For all other low-temperature structures as well, we have obtained the
 optimal values of the lattice parameters and atomic positions, starting
 from the available experimental values
 \cite{TauberAntimonyInterest5,Faik2012}. The structures at 613K, 430K
 and room temperatures were considered, which, respectively correspond
 to $I4/m$, $I2/m$ and $P2_1/n$ space groups. Additionally, for $I2/m$
 phase, the experimental lattice parameters and atomic positions were
 first transformed to the $B2/m$ representation as this is the only
 representation supported by WIEN2k for space group No. $12$.
 The optimal values of the lattice parameters correspond to the atomic positions such that the average force per atom is less that 1 mRy/a.u.. The calculated results are then fitted using the Birch-Murnaghan equation of state (BM-EOS) \cite{Birch}:

 \begin{equation}
    \small
    E(V)=E_0 + \frac{9V_0B_0}{16} \left\{ \left[ \left( \frac{V_0}{V} \right )^{2/3} -1 \right]^3 B'_0 +        \left[ \left( \frac{V_0}{V} \right )^{2/3} -1 \right]^2 \left[6 - 4 \left( \frac{V_0}{V} \right         )^{2/3} \right ] \right\}
    \label{eqn:birch}
  \end{equation}
 where, $V_0$ and $B_0 (= -V (\partial P/ \partial V)_T)$ are the equilibrium
 volume and bulk modulus, respectively.
 $B'_0$ is the pressure derivative of $B_0$ and is, like $B_0$, evaluated at
 $V_0$. The unit cell parameters and atomic positions used in the calculations is presented
 in the Supplementary Material.

 \begin{table}
 \centering
 \begin{tabular}{c| c| c| c| c| c|} \hline
      \multicolumn{1}{|c|}{\multirow{2}{*} {Atom}} &
      \multicolumn{4}{c|}{SSS} &
      \multicolumn{1}{c|}{CSS} \\ \cline{2-6}
      \multicolumn{1} {|c|} {} &
      \multicolumn{1} {c|}{$P2_1/n$} &
      \multicolumn{1} {c|}{$B2/m$} &
      \multicolumn{1} {c|}{$I4/m$} &
      \multicolumn{1} {c|}{$Fm \bar3m$} &
      \multicolumn{1} {c|}{$P2_1/n$} \\ \cline{1-6}


      \multicolumn{1} {|c|} {Sr/Ca} &
      \multicolumn{1} {c|}{2.39} &
      \multicolumn{1} {c|}{2.45} &
      \multicolumn{1} {c|}{2.47} &
      \multicolumn{1} {c|}{2.5} &
      \multicolumn{1} {c|}{2.21} \\

      \multicolumn{1} {|c|} {Sb} &
      \multicolumn{1} {c|}{1.89} &
      \multicolumn{1} {c|}{1.97} &
      \multicolumn{1} {c|}{1.88} &
      \multicolumn{1} {c|}{1.87} &
      \multicolumn{1} {c|}{1.89} \\

      \multicolumn{1} {|c|} {Sc} &
      \multicolumn{1} {c|}{2.10} &
      \multicolumn{1} {c|}{1.99} &
      \multicolumn{1} {c|}{2.06} &
      \multicolumn{1} {c|}{2.07} &
      \multicolumn{1} {c|}{2.10} \\ 

      \multicolumn{1} {|c|} {O} &
      \multicolumn{1} {c|}{1.86} &
      \multicolumn{1} {c|}{1.76} &
      \multicolumn{1} {c|}{1.83} &
      \multicolumn{1} {c|}{1.84} &
      \multicolumn{1} {c|}{1.86} \\ \cline{1-6}

 \end{tabular}
      \caption{Values of the Muffin-tin radii, $R_{\rm MT}$, for different
      phases of SSS and
      CSS. The values of $R_{\rm MT}$ are expressed in ${\rm \AA}$.}
      \label{table:MTValues}
 \end{table}

 Table \ref{table:MTValues} lists the values of muffin-tin radii, $R_{\rm
 MT}$, for
 different atoms corresponding to different crystal structures used for
 the final calculations. For volume optimization studies, $10\%$
 reduction in $R_{\rm MT}$ values were used.
 In the following 
 calculations, the values of $R \times k_{\rm max}$ was $7.0$, where
 $k_{\rm max}$ is the plane-wave cut-off and $R$ is the smallest of all
 ionic radii. We used approximately 2000 $k$-points in the
 Brillouin zone for numerically performing the integration over $k$,
 using the tetrahedron method, while calculating the DOS and
 bandstructure. While the core states were
 treated relativistically, the semi-core and valence states are treated
 semi-relativistically, \textit{i.e.}, ignoring the spin-orbit coupling.
 The cut-off energy, which defines the separation between the core and
 valence states was appropriately chosen to be -8.0 Ry.
 The density of states and bandstructure is found to
 be spin-independent due to the empty or fully filled shells of the Sc
 and Sb atoms, as discussed in the following. The exchange and
 correlation effects have been treated within the standard
 parametrization \cite{pbe} of the Generalized Gradient
 Approximation (GGA). The self-consistency is better than $0.001 {\rm e/a.u.^3}$ for
 the charge density and the stability is better than $0.01 {\rm mRy}$ for the
 total energy per unit cell. 

 The complex dielectric tensor was calculated in the WIEN2k program
 using the well-known relation \cite{Optical}:
 \begin{eqnarray}
    \epsilon_2(\omega)={\rm Im}\, \epsilon_{ij}(\omega)&=& \frac{4\pi}{m^2
        \omega ^2} \int d{\bf k} \sum_{nn'}\langle {\bf k}n \mid {\bf p}_i \mid {\bf k}n'\rangle \langle {\bf k}n' \mid {\bf p}_j \mid {\bf k}n \rangle \nonumber \\
    & \times & \delta (E_{{\bf k}n} -E_{{\bf k}n'}
 - \hbar \omega)
  \,\,,\,\,\,\,
 \label{eqn:imageps}
 \end{eqnarray}
 where, $i,j=(x,y,z)$ are the three Cartesian directions, ${\bf p}_i
 \equiv (\hbar/i) \nabla_i$ is
 the momentum operator along the direction $i$, $\mid {\bf k}n \rangle$
 is a crystal wavefunction, and $\omega$ is the photon
 energy (in the units of $\hbar =1$). The $\delta$-function is
 approximated as a Lorentzian with width $\Gamma =
 0.1$ eV for the calculations presented in this work. 
 The real part of the dielectric
 function is calculated using the Kramers-Kronig relation:
 \begin{equation}
 \epsilon_1(\omega)={\rm Re}\, \epsilon_{ij}(\omega)= \delta_{ij} +
 \frac{2}{\pi} P \int_0 ^\infty \frac{\omega'\,\, {\rm Im}\: \epsilon_{ij}
 (\omega ')}{\omega^{'2} -\omega ^2} d\omega'\,\,,
 \label{eqn:reepskkr}
 \end{equation}
 and the optical conductivity is given by:
 \begin{equation}
 \sigma_{ij} (\omega) = -i\:\frac{\omega}{4\pi} \epsilon_{ij} (\omega)\,.
 \label{eqn:sigmak}
 \end{equation}

 Optical constants, such as the refractive index $n(\omega)$, and the
 extinction coefficient $k(\omega)$ are calculated in terms of the real
 and imaginary parts of the dielectric response as: 
 \begin{eqnarray}
      n(\omega) &=& \frac{1}{\sqrt{2}} \bigg[ \sqrt{\epsilon_1^2 (\omega) +
      \epsilon_2^2 (\omega)} + \epsilon_1 (\omega) \bigg]^{1/2} \nonumber \\
      k(\omega)& =& \frac{1}{\sqrt{2}}\bigg[ \sqrt{\epsilon_1^2 (\omega) +
      \epsilon_2^2 (\omega)} - \epsilon_1 (\omega) \bigg]^{1/2} \,,
 \label{eqn:nandk}
 \end{eqnarray}
 which leads to the following relation in the low-frequency limit:
 \begin{equation}
      n(0)=\sqrt{|\epsilon(0)|}\,,
      \label{eqn:staticdielectric}
 \end{equation}
 where $\epsilon(0)$ is the complex dielectric constant in the
 corresponding limit. 

 The absorption coefficient $\alpha(\omega)$, reflectance
 $R(\omega)$, and
 the energy-loss spectrum $L(\omega)$ may also be obtained in terms of
 the real and imaginary part of the dielectric function using simple
 mathematical relations: 
 \begin{eqnarray}
      \alpha(\omega) &=& \sqrt{2} \omega \bigg[ \sqrt{\epsilon_1^2 (\omega) +
      \epsilon_2^2(\omega)} - \epsilon_1 (\omega) \bigg]^{1/2} \nonumber \\
      R(\omega) &=&\left| \frac{\sqrt{|\epsilon(\omega)}| -1}
      {\sqrt{|\epsilon(\omega)|} +1} \right|^2= \frac{(n(\omega)-1)^2 +
      k^2(\omega)}{(n(\omega)+1)^2
 + k^2(\omega)} \nonumber \\
      L(\omega) &=& \frac{\epsilon_2 (\omega)}{ \bigg[
            \sqrt{\epsilon_1^2 (\omega) +
      \epsilon_2^2 (\omega)} - \epsilon_1 (\omega) \bigg]^{1/2}} \,.
 \label{eqn:absandlosscoeff}
 \end{eqnarray}

\section{Results and Discussion}

\subsection{Kubelka-Munk Spectrum and Optical Bandgap}
 Fig \ref{fig:VolOpt}(a) shows the KM spectrum as a function of the
 incident photon energy for the room temperature phases of SSS and
 CSS for $n=1/2$ corresponding to the direct bandgap for these
 compounds. Presence of a direct bandgap is consistent with 
 the DFT bandstructure calculations discussed
 in the following subsections. Interestingly, the difference in the
 bandgap values for these compounds is very small. The bandgap
 values for SSS and CSS is found to be approximately 3.57 eV and
 3.68 eV, respectively, 
 A difference of 0.11 eV due to smaller A-site cation is consistent with earlier studies on other ${\rm Sc}$ based
 $d^0$ double perovskites and is captured well by DFT calculation, as
 will be discussed later. In other ${\rm Sc}$ based $d^0$
 double perovskite systems, such as ${\rm Sr_2ScNbO_6}$ and ${\rm Sr_2ScTaO_6}$, 
 changing the A-site cation from ${\rm
 Sr}$ to ${\rm Ca}$ leads to increase in bandgap values as large as
 0.33 eV \cite{Eng2003}. The relative ineffectiveness of the A-site cation in inducing large bandgap
 changes is due to large difference 
 in electronegativity values of the ${\rm B}$ and ${\rm B'}$-site
 transition metals, as discussed in the following. 

\subsection{Structural Optimization}

 \begin{table}[ht]
 \centering
 \begin{tabular}{|l| c| c| c|c|}
 \hline  
 Sample, Space group & $V_0 ({\rm a.u.}^3/{\rm f.u})$ & $B_0 ({\rm GPa})$ & $B_0'$ &
 $E_0 ({\rm Ry})$\\ [1.0ex]
 \hline 
 SSS, $Fm \bar{3}m$ & 910.90 & 143.3 & 5.2 & -28119.09\\ 
 SSS, $I4/m$ & 906.75 & 140.0 & 4.9 & -28119.17 \\ 
 SSS, $B2/m$ & 909.56 & 139.3 & 4.7 & -28119.11 \\ 
 SSS, $P2_1/n$ & 905.96 & 141.9 & 4.7 & -28119.11\\ 
 CSS, $P2_1/n$ & 846.63 & 142.3 & 4.8 & -18121.30 \\
 \hline
 \end{tabular}
      \caption{Values of the parameters $V_0$, $B_0$,
      $B_0'$, and the minimum energy $E_0$ for the Birch-Murnaghan equation of
      state for the different phases of the SSS and CSS compound. Please refer
      text for details.}
      \label{table:bmeos}
 \end{table}

\begin{figure}
 \begin{center}
       \includegraphics[angle=-90,scale=0.165]{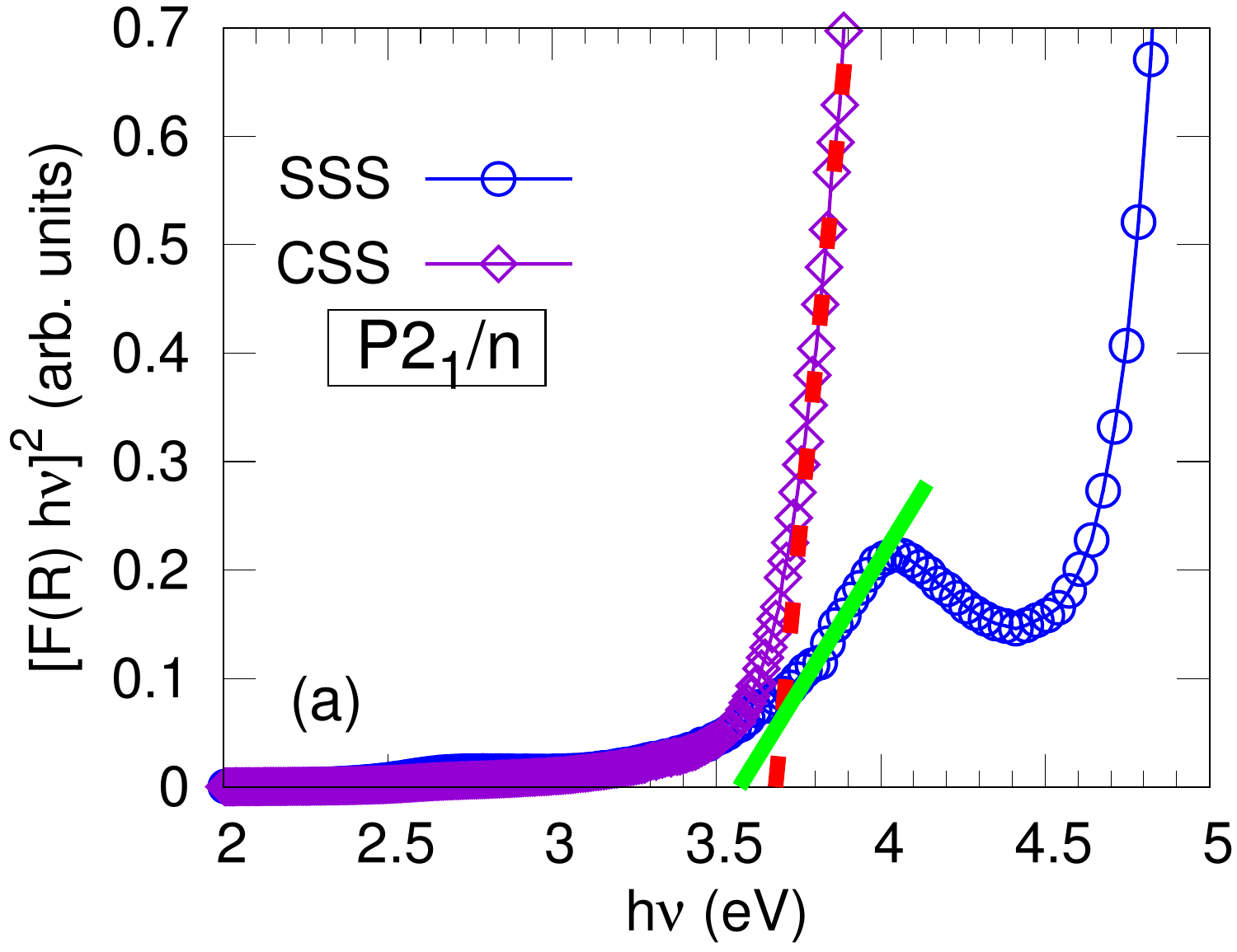}
       \includegraphics[angle=-90,scale=0.182]{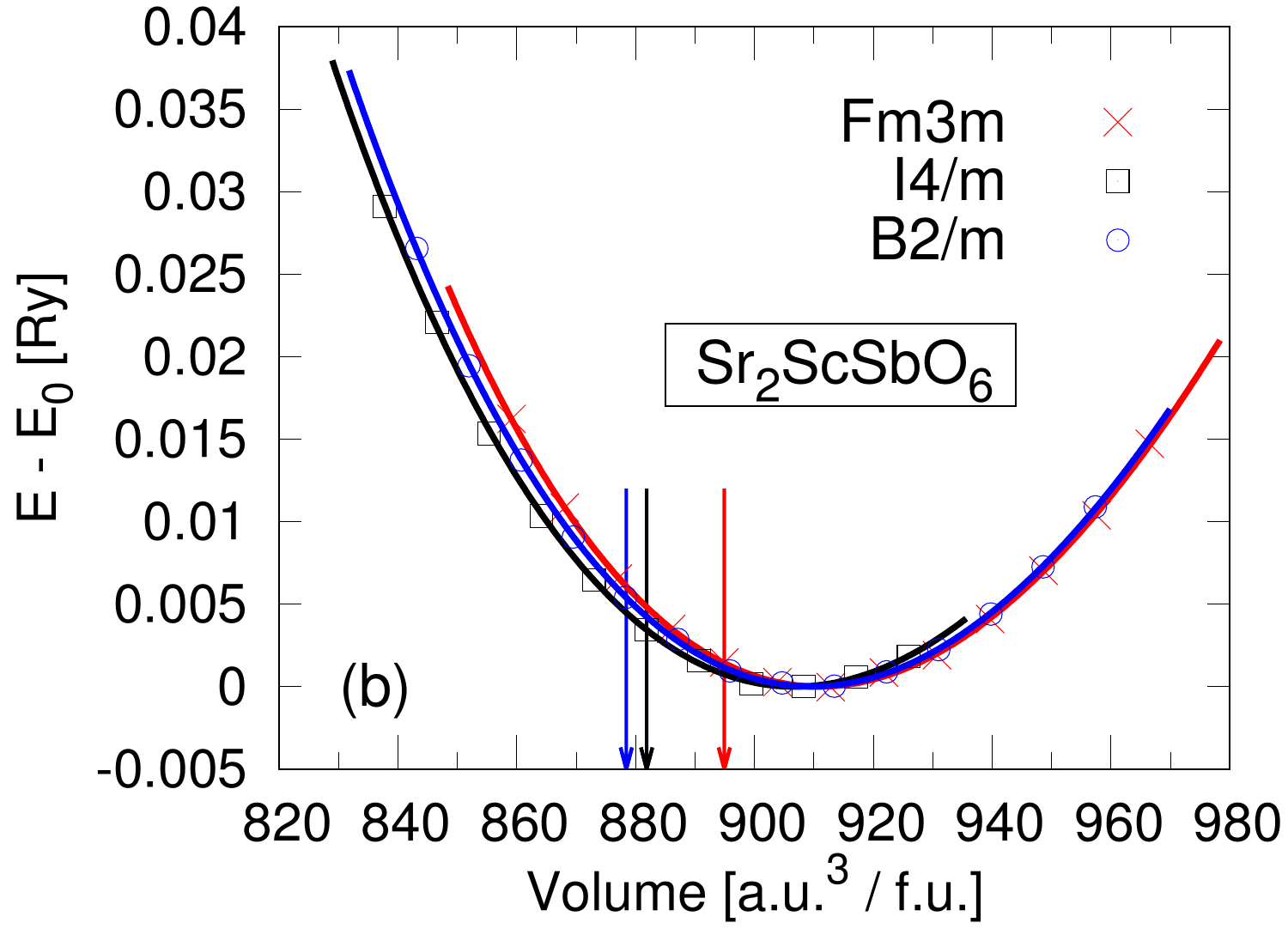}
       \includegraphics[angle=-90,scale=0.182]{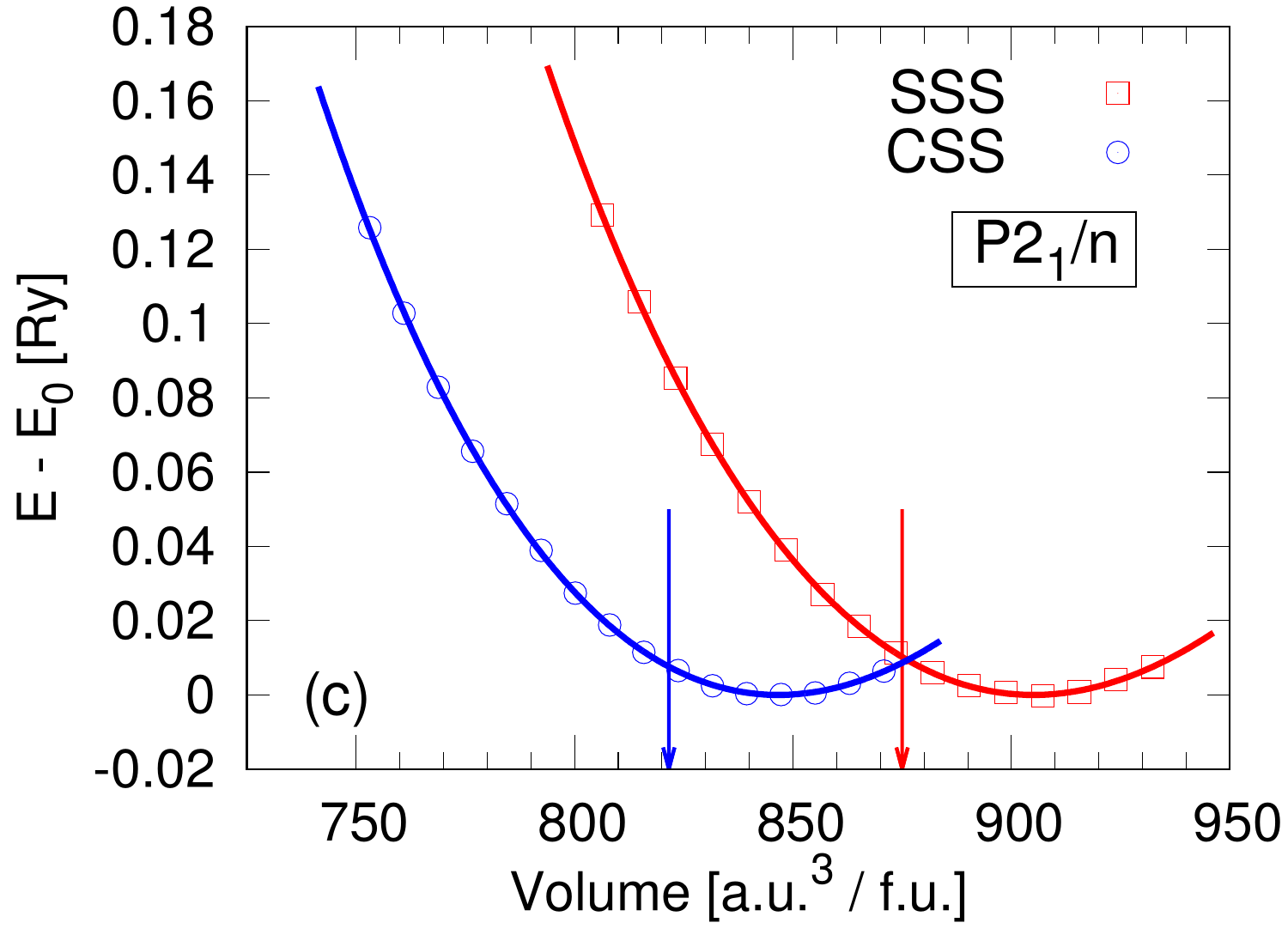} \\
 \end{center}
       \caption{Color Online. (a) Kubelka-Munk function as a function of energy for
       the room temperature phases of SSS (circles) and CSS (diamonds).
       The solid green (dashed red)
       line is the extrapolation of the straight portion
       of the curve at the absorption edge for CSS (SSS). 
       (b)-(c) Energy as a function of the unit cell volume per formula unit
       (f.u.) for different
       phases of the SSS and CSS. The solid curve is a fit to the
       corresponding Birch-Murnaghan equation of state (Eq. (\ref{eqn:birch})).
       The arrows indicate the corresponding experimental
       values. The energies have been shifted for comparison. Values of $E_0$ are provided in Table \ref{table:bmeos}
       }
      \label{fig:VolOpt}
 \end{figure}

 DFT calculations were performed with the relaxed structural parameters for all the phases of these compounds. Fig. \ref{fig:VolOpt}(b)-(c) shows the total energy variation as a function of the unit cell volume per formula unit (f.u.) for different phases of the SSS and CSS compounds (data points). The smooth curves are the corresponding fit to the BM-EOS (Eq. (\ref{eqn:birch})), allowing us to obtain the equilibrium structural parameters for each phase.
 The optimal unit cell volumes, bond lengths and angles are found to be in reasonable agreement with corresponding experimental values (see Supplementary Material for details).
 Table 2 shows the equilibrium values of the parameters in the BM-EOS, \textit{viz.},  $V_0$, $B_0$, $B_0'$ and $E_0$, described earlier.

\subsection{Electronic Properties}
 We begin with the observation that in all the cases, there is
 hybridization between the O-$p$ and Sc-$d$, Sb-$s$ orbitals, leading to
 a valence band (VB) with states of dominantly O-$2p$ non-bonding
 character and the low lying conduction band (CB) with states of
 O-$2p$, Sb-$5s$ and Sc-$3d$ character. However, the nature of
 hybridizing orbitals depends on the crystal structure as discussed
 below. We start with the most symmetric (ideal), high temperature cubic phase.

 \begin{figure}
 \begin{center}
       \includegraphics[angle=-90,scale=0.0925]{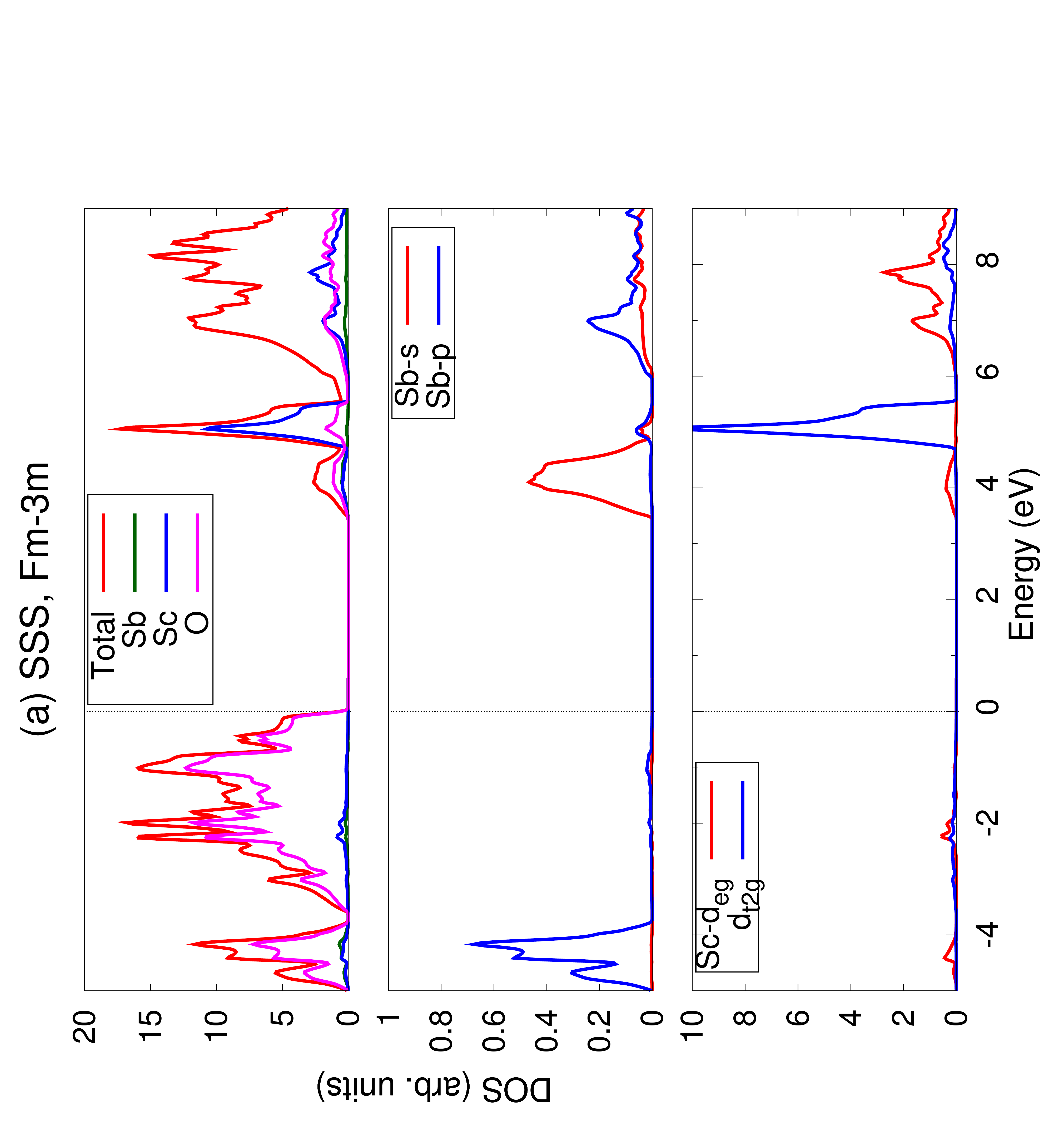}
       \includegraphics[angle=-90,scale=0.0925]{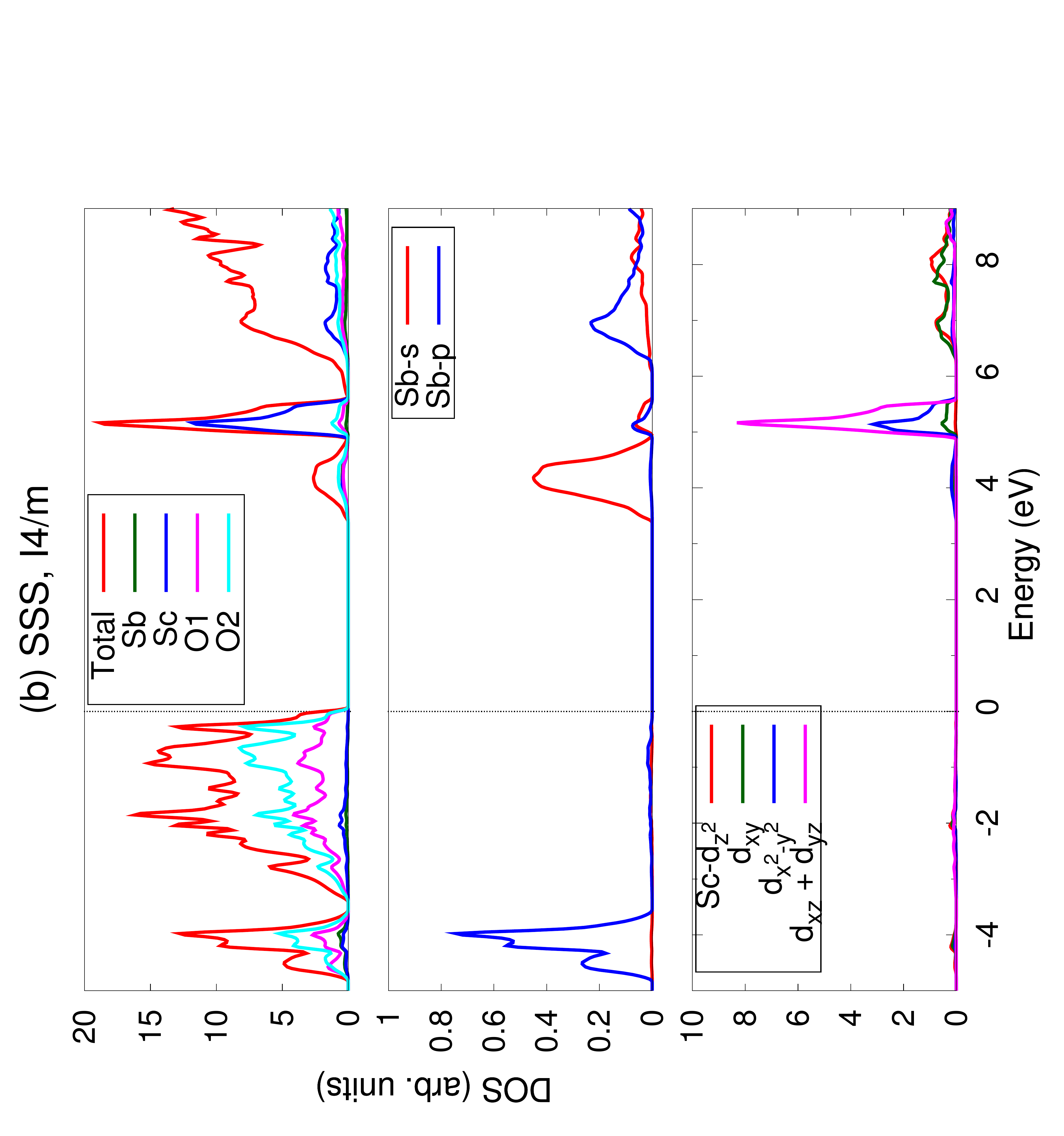}
       \includegraphics[angle=-90,scale=0.0925]{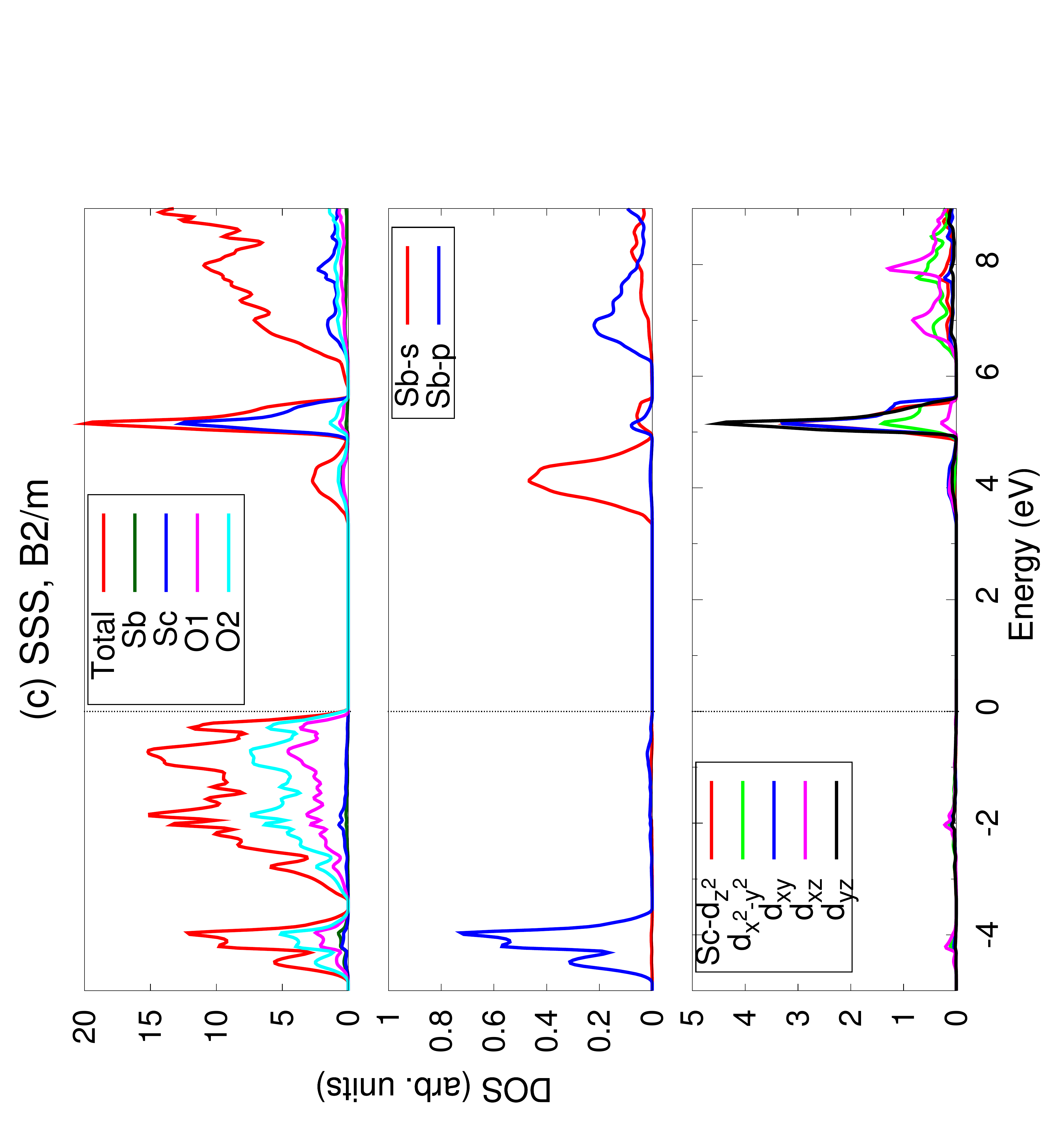}
       \includegraphics[angle=-90,scale=0.0925]{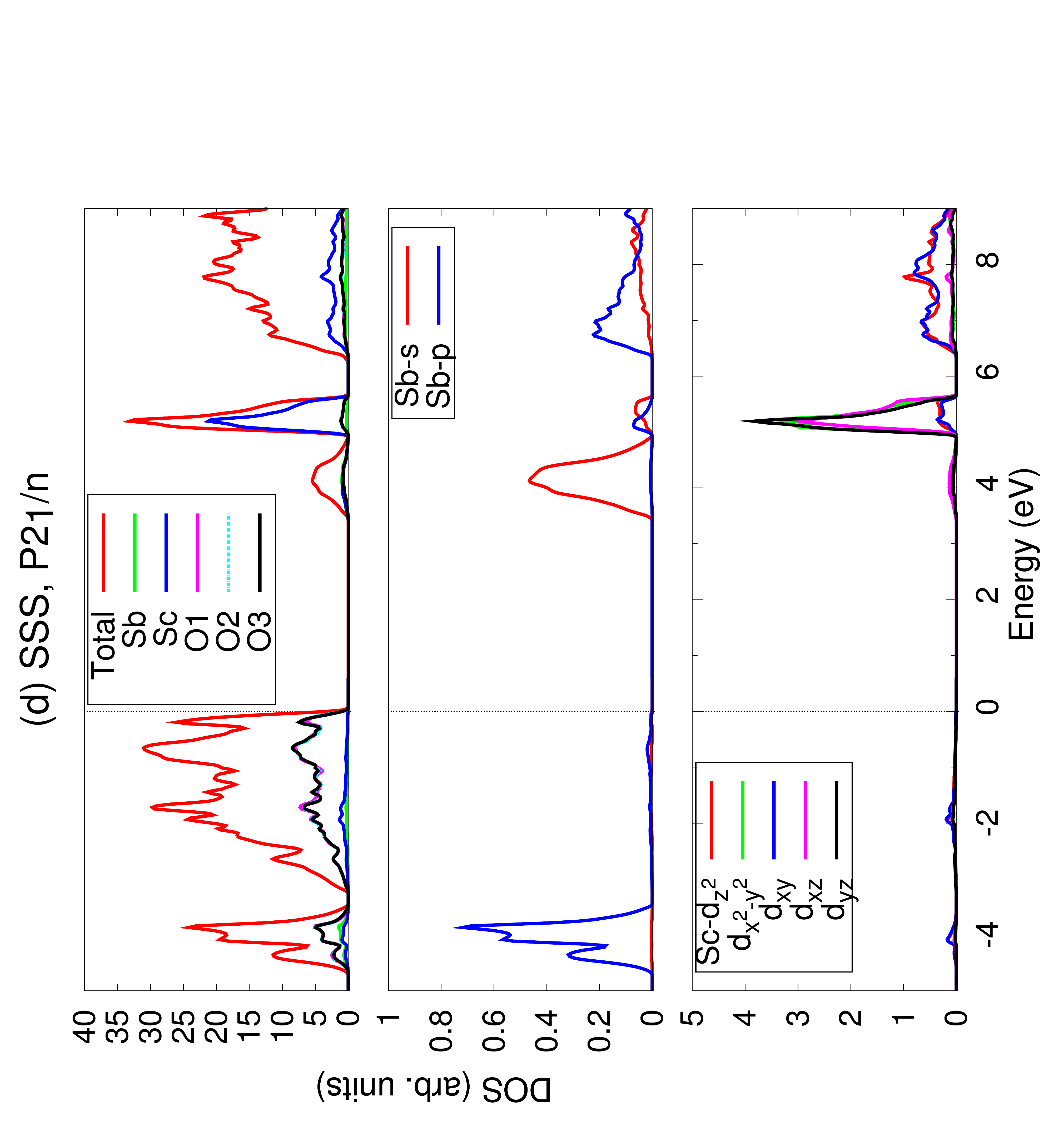}
 \end{center}
      \caption{Color Online. Total and partial density of states (DOS) per f.u. 
       for different phases of SSS. Note the different scaling of the DOS-axes.
      }
      \label{fig:dos}
 \end{figure}

\subsubsection{Cubic High-temperature Structure}

\begin{figure}
 \begin{center}
       \includegraphics[angle=0,scale=0.470]{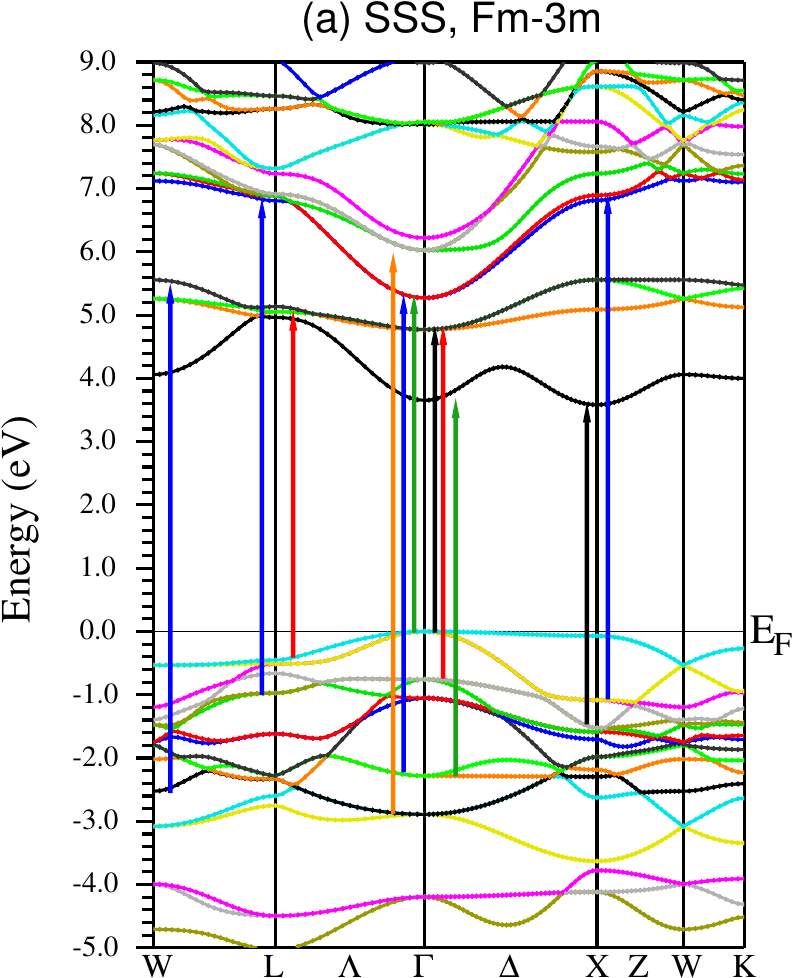} 
       \includegraphics[angle=-0,scale=0.470]{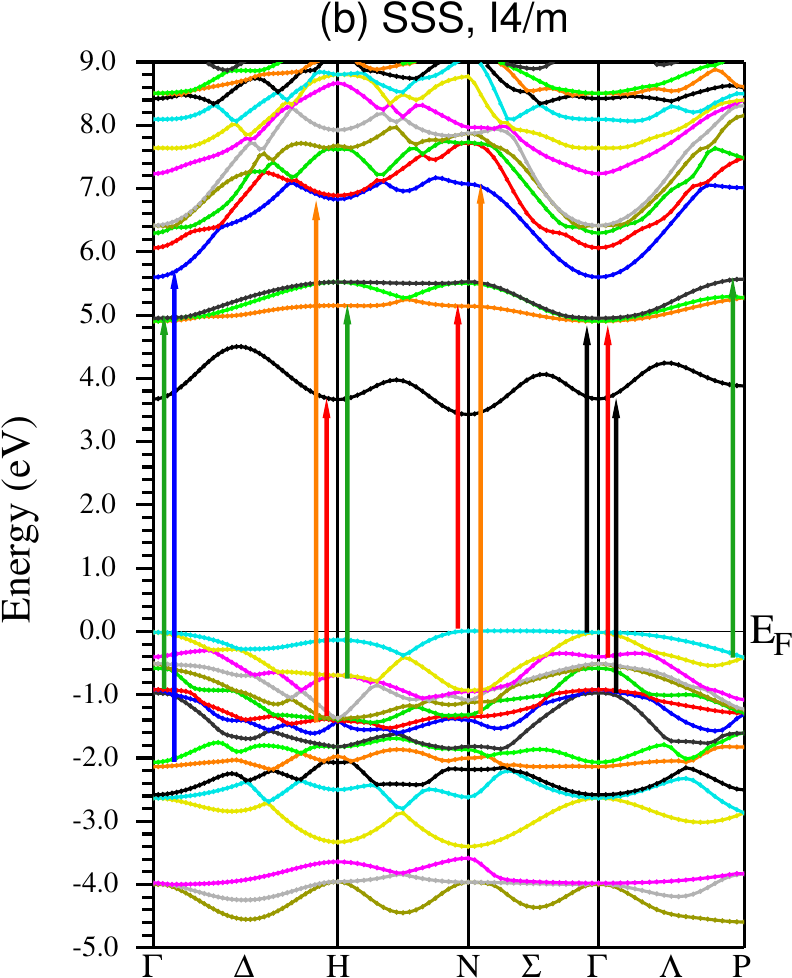}
       \includegraphics[angle=-0,scale=0.470]{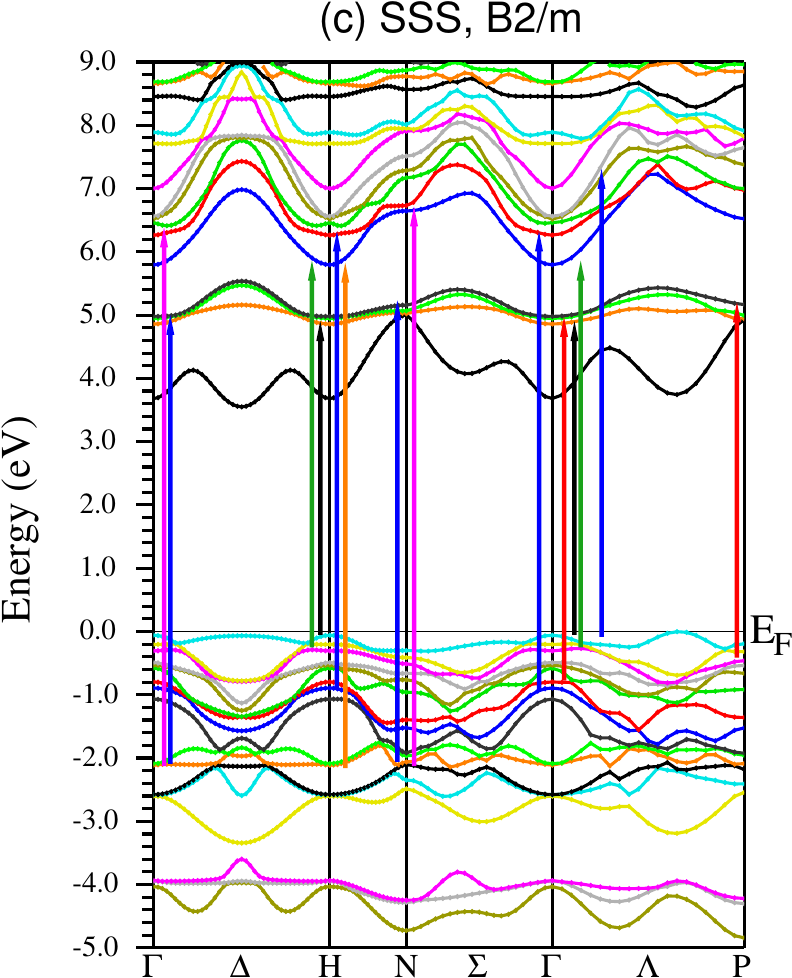}
       \includegraphics[angle=-0,scale=0.470]{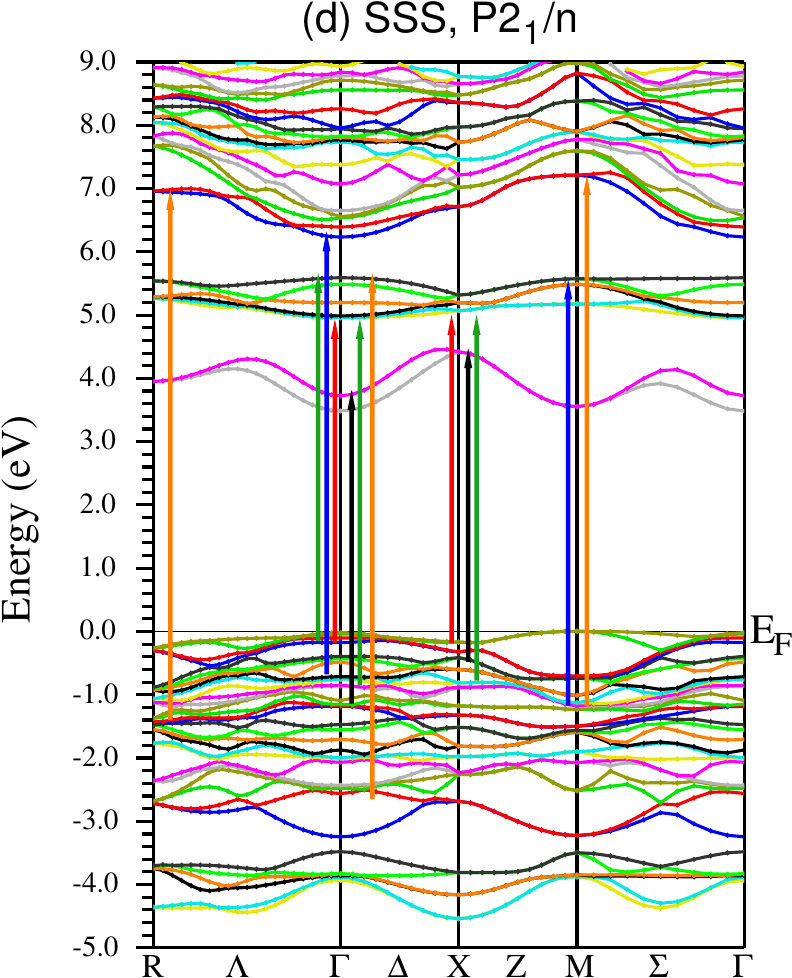}
 \end{center}
      \caption{Color Online.  Bandstructure for the different phases
      of the SSS. The arrows indicate possible optical
      transitions corresponding to different peaks in Fig.
      \ref{fig:dielectric}: black for peak A, red for B, green for C,
      blue for D, orange for E and pink for F.
      }
      \label{fig:bandstr}
 \end{figure}

 Fig. \ref{fig:dos}(a) shows the total and partial density of states (DOS) for the cubic phase at $660$ K. The electronic bandgap is found to be $3.51$ eV. The dominant contribution to the CB comes from the O-$2p$ and the Sb-$5s$ orbitals, forming the lowest unoccupied state, and the Sc-$3d$
 orbitals. Since the Sc atom is located at the center of an octahedron, the $d_{x^2-y^2}$ and $d_{z^2}$ orbitals forming the $e_{\rm g}$ band point towards the neighboring O atoms, whereas the $d_{xy}$, $d_{yz}$, and $d_{xz}$ orbitals forming the $t_{\rm 2g}$ band point between the O-$2p$ orbitals. 
 The crystal field due to the O atoms lifts the degeneracy of the Sc-$d$ states, breaking them into
 $t_{\rm 2g}$ and $e_{\rm g}$ bands, with $t_{\rm 2g}$ states lying lower in energy. The approximate bandwidths of $t_{\rm 2g}$ and $e_{\rm g}$ bands is approximately $0.8$ and $3.0$ eV, respectively. Fig. \ref{fig:bandstr}(a) shows the corresponding bandstructure. The crystal field splitting between the $t_{\rm 2g}$ and $e_{\rm g}$ band is approximately $1.2$ eV at the $\Gamma$ point.

 For the Sb atom, both the $s$- and a major part of the $p$-orbitals lie in the CB reflecting the +3 valency. While the $s$ states lie at the CB edge, the $p$ states, hybridized with the O-$p$ orbitals, lie relatively higher into the CB, closer to the Sc-$d_{\rm eg}$ states. Although both ${\rm Sc}$ and ${\rm Sb}$ contribute to the CB at low energies, there is little mixing between the ${\rm Sc}$-$d$ states with the ${\rm Sb-O}$ orbitals due to large electronegativity difference between the two. With contribution from Sb-$p$, the states in the crystal field gap are dominantly of O-$p$ character.
 Contributions from the Sr atom lie high in the CB (not shown). 

\subsubsection{Distorted Low-temperature Structures}
 Fig. \ref{fig:dos}(b)-(d) and Fig. \ref{fig:bandstr}(b)-(d), respectively, show the DOS and the electronic bandstructure of SSS in $I4/m$ (at $613$ K), $B2/m$ (at $430$ K) and $P2_1/n$ (at room temperature) phases. The broad features of CB and VB
 remain qualitatively similar in the lower symmetry phases. The CB and VB edges, respectively, still contain states with Sb-$s$ and O-$p$ character. However, with increasing octahedral distortions, it is expected that the nature of hybridization between Sc-$d$ and O-$p$ orbitals will change. Indeed, we find that the Sc-$d$ orbitals participating in the formation of effective $t_{\rm 2g}$ band changes according to the octahedral tilt and rotation of the
 distorted unit cell with respect to the cubic phase. For example, the tetragonal unit cell can be obtained from the cubic unit cell by a $45^\circ$ rotation about $z$-axis. This, in turn, interchanges the $d_{xy}$ and $d_{x^2-y^2}$ orbitals. Also, $d_{xz}$ and $d_{yz}$ orbitals are symmetric with respect to the position of O atoms at the $\Gamma$ point. Therefore, in this phase, the effective $t_{\rm 2g}$ band is formed out of states with $d_{xz}$, $d_{yz}$ and $d_{x^2-y^2}$ character, while these states remain quasi-degenerate at the $\Gamma$ point. 
 
 Further increase in the octahedral distortion breaks the symmetry of the $d_{xz}$ and $d_{yz}$ orbitals. Reduced symmetry in $B2/m$ phase leads to mixing of different characters in states forming the $t_{\rm 2g}$ band. This is accompanied with changes in the dispersion of non-bonding O-$p$ states in the VB edge, from being relatively dispersion-less to quadratic around the $\Gamma$ point, which could be important for hole doping in these materials. For the monoclinic $P2_1/n$ phase, on the other hand, composition of the $t_{\rm 2g}$ band is similar to the tetragonal phase. However, there are three inequivalent O atoms per unit cell. The number of states in this phase is doubled due to doubling of the unit cell. Nevertheless, there is almost no change in the effective $t_{\rm 2g}$ bandwidth in lower symmetry phases. 
 
 The absence of changes in $t_{\rm 2g}$ band is further reflected in minimal changes in the electronic bandgap values. The bandgap values for the tetragonal $I4/m$, monoclinic $B2/m$ and monoclinic $P2_1/n$ phases are found to be 3.41, 3.40 and 3.47 eV respectively. Therefore, as compared to the cubic phase, the band gap values first decrease and then increase with increasing distortion of the octahedra. The maximum change in bandgap values of different phases of SSS is approximately 0.1 eV. Also, the bandgap values for the room temperature phase is in a reasonably good agreement with the corresponding experimental value of $3.57 {\rm eV}$ obtained in this work using the diffuse reflectance spectroscopy (see Figure \ref{fig:VolOpt}(a), Section 3.1), as it is known that DFT calculation underestimate the bandgaps.

 The most prominent feature at lower temperatures is that, compared to the cubic phase, there is very little change in the distorted phases. This is consistent with the small change in the angle $\beta$ in distorted monoclinic $B2/m$ and $P2_1/n$ phases of the SSS compound (see Supplementary Material for details). Furthermore, the bandgap values obtained in this work are qualitatively consistent with earlier studies on other Scandium based $d^0$ double perovskites, ${\rm A}_2{\rm ScB'O}_6$ (${\rm A}={\rm Sr,Ca}$; ${\rm B'=Ta, Nb}$) \cite{Eng2003}. As Sb is more electronegative than both Ta and Nb, it would be expected that the electronic bandgap of SSS and CSS be lower than that of Ta and Nb-based compounds. In Ta and Nb-based compounds, such lowering of band gap is accompanied by a change in $t_{2g}$ bandwidth induced by hybridization between ${\rm Sc}$-$d$ and ${\rm (Ta/Nb)-O}$ states. In contrast, in
 the present case, due to large difference in the electronegativity values of ${\rm Sc}$ and ${\rm Sb}$, the effective $t_{\rm 2g}$ bandwidth remains unchanged. This effect arises out of the non-mixing
 of ${\rm Sc}$ and ${\rm Sb-O}$ states in the CB. Therefore, in the present case, the effects of increase in octahedral tilting, without change in electronegativity and other parameters, in $d^0$ double perovskite systems is minimal. 

 \begin{figure}
 \begin{center}
       \includegraphics[angle=0,scale=0.065]{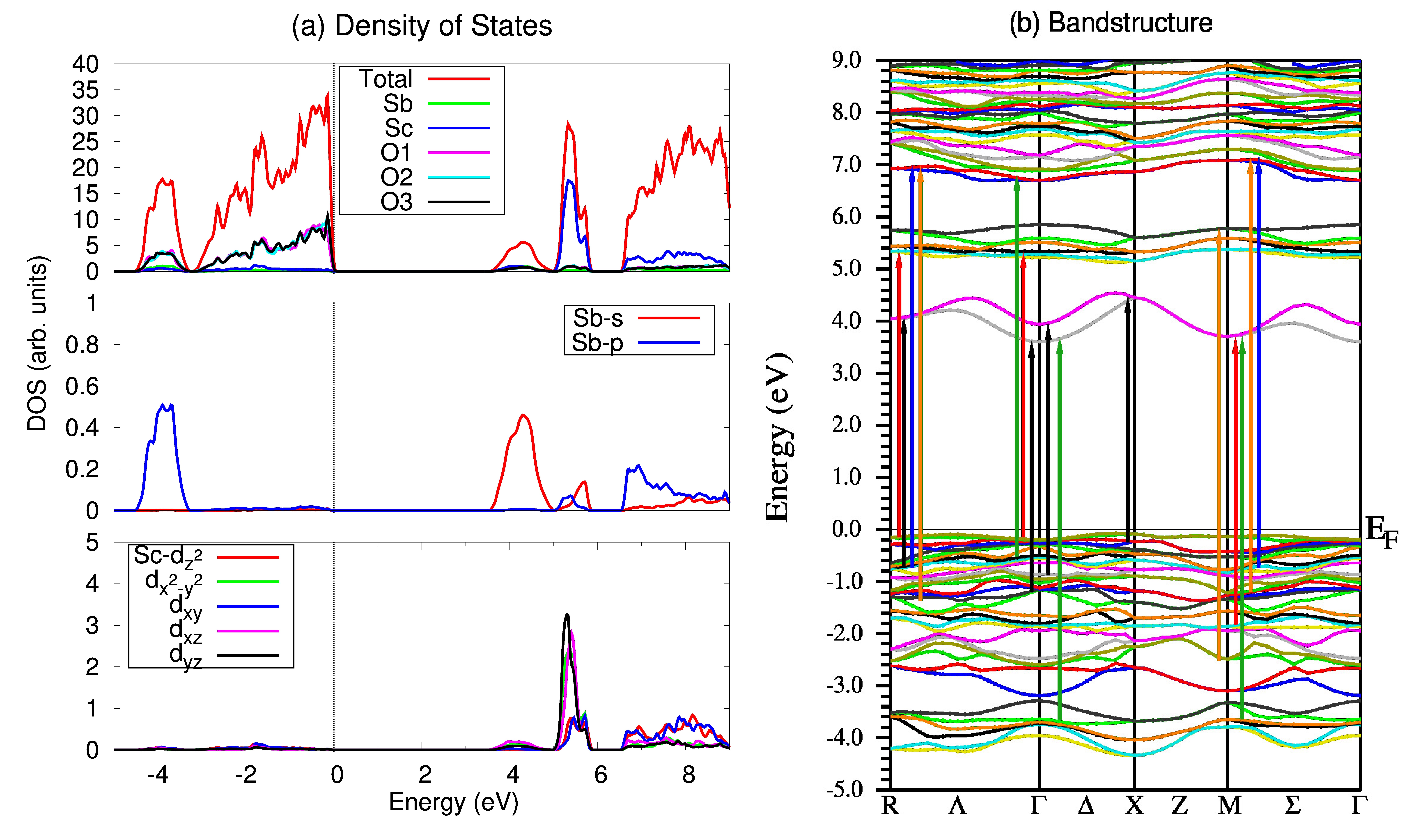}
 \end{center}
      \caption{Color Online. DOS  and bandstructure for the room temperature,
      monoclinic $P2_1/n$ phase of the CSS. The arrows indicate optical
      transitions corresponding to different peaks in Fig.
      \ref{fig:dielectric}: black for peak A, red for B, green for C,
      blue for D, and orange for E.
      }
      \label{fig:css}
 \end{figure}

 A comparison with the DOS and bandstructure of the CSS, shown in Fig. \ref{fig:css}, shows that the room temperature phases of the two compounds are remarkably similar with respect to the Sc, Sb and O contributions, which play important role in electrical, transport and optical properties as they lie close to the band edge. However, a comparison of the total DOS shows that the CSS compound has a larger bandgap, approximately equal to $3.61$ eV. This is again in good agreement with the corresponding experimental value of $3.68 {\rm eV}$ as discussed in Section 3.1 (see Figure \ref{fig:VolOpt}(a)). 
 Yet another difference is the dispersion of the states in the VB lying close to the band edge. While for SSS, the band edge occurs at the M point with a relatively dispersion-less state, the corresponding state in the CSS are dispersive around the band edge at the X point. This would affect the Fermi surface and localization of charges on hole doping. Furthermore, there are also changes in the dispersion of the relatively deep VB O states, Although the
 difference is not large, they lead to significant changes in the optical properties of the room temperature phases of these compounds. These differences can be attributed to the relative smallness of the Ca cation. Although both Sr and Ca have the valency of +2, the relative smallness of the Ca atoms
 leads to a larger octahedral distortion and, therefore, larger bandgap for the CSS compound. The relative
 difference between the ionic radii of Sr and Ca is $\Delta r = 1.32{\rm \AA} -1.14{\rm \AA} = 0.18{\rm \AA}$.

\subsection{Finite-frequency Optical Properties}
 Optical properties are studied through the complex dielectric function $\epsilon (\omega)$ which is a measure of the linear response of the system to an external electromagnetic field:
 \begin{equation}
       \epsilon(\omega)=\epsilon _1(\omega) + i \epsilon _2 (\omega)\,,
 \end{equation}
 where $\epsilon_1$ and $\epsilon_2$ are, respectively, the real and imaginary parts of the dielectric response and are connected through the Kramers-Kronig relations (Eq. (\ref{eqn:reepskkr})). The values of $\epsilon_1 (\omega)$ and $\epsilon_2(\omega)$ are obtained thorough the WIEN2k program, as described
 earlier in Eq. (\ref{eqn:imageps}) and (\ref{eqn:reepskkr}).

\subsubsection{Dielectric Response Function}
 In general, the dielectric response is a rank-2 tensor with a total of nine variables. However, the
 structure of this tensor greatly simplifies due to the symmetries present in the crystal structures, often leading to separation of $x$, $y$ and $z$ components of the dielectric tensor.
 While for cubic symmetry, there is only one independent component, for tetragonal and monoclinic geometry, it would suffice to study only the $x$ ($\epsilon_{xx}$) and the $z$ components ($\epsilon_{zz}$). In the following, we will study the optical properties through the averaged response in the $x$ and $z$ directions, unless mentioned otherwise.

\begin{figure}
 \includegraphics[angle=-90,scale=0.105]{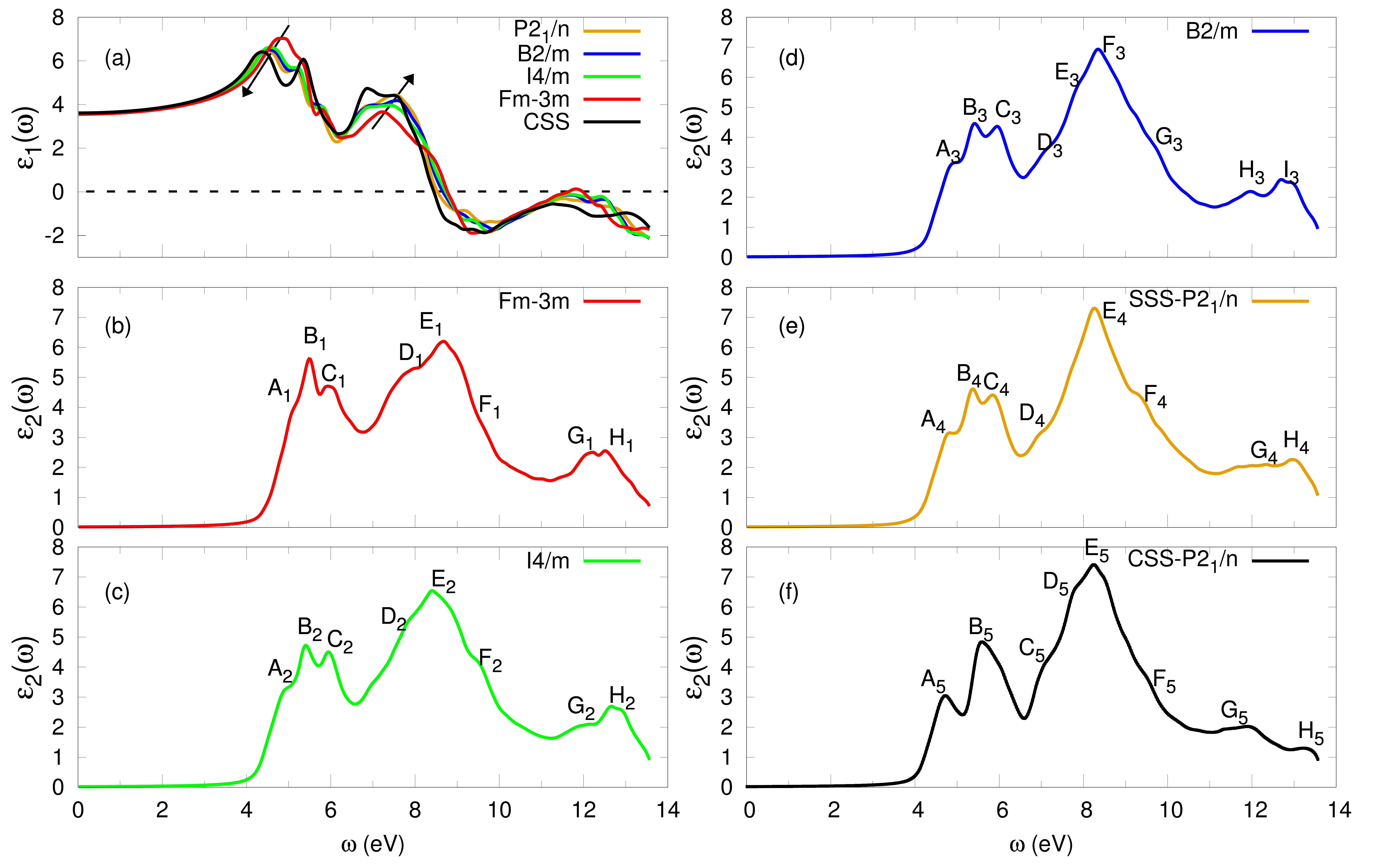}
 \caption{Color Online. Real $\epsilon_1(\omega)$ and imaginary $\epsilon_2(\omega)$
       parts of the linear dielectric response for the
       different phases of the CSS and SSS.}
 \label{fig:dielectric}
 \end{figure}

 Fig. \ref{fig:dielectric} shows the photon energy (frequency) dependence of the inter-band dielectric constants for the different phases of the CSS and SSS compounds. Fig. \ref{fig:dielectric}(a) shows the dispersive (real) part of the dielectric response $\epsilon_1(\omega)$ as a function of the photon energy. Presence of multiple peaks in $\epsilon_1(\omega)$ is due to the multiple photon processes. Interestingly, the position of the first (second) peak decreases (increases) continuously as the crystal structure changes from cubic to monoclinic, as shown by the arrow in the plots. At the same time, the adjoining shoulder on the right (left) of the first (second) peak evolves into a well-resolved sharper peak, resulting in a total of four low-energy peaks. At high frequencies, the zero-crossing frequency of $\epsilon_1(\omega)$, which corresponds to the location of the screened plasma frequency, decreases monotonically with decreasing temperature, occurring at approximately $8.77$ eV for the $Fm \bar{3}m$, $8.73$ eV for the $I4/m$, $8.66$ for the $B2/m$, and $8.53$ eV for the $P2_1/n$ phase of the SSS and
 $8.45$ eV for the CSS.

 The imaginary part of the dielectric response, $\epsilon_2(\omega)$, closely follows the DOS and
 starts from a finite frequency due to the finite electronic bandgap in these compounds. The peaks observed in the $\epsilon_2(\omega)$ at finite frequencies originate from the transitions between states across
 the Fermi energy. The characteristic peaks for different phases of the SSS and CSS compounds
 have been labeled (Fig. \ref{fig:dielectric}, Panels (b)-(f)). As can be seen from Fig. \ref{fig:bandstr} and \ref{fig:css}, it is possible to assign multiple transitions corresponding to the peak energy values. This is done by identifying the states across the Fermi energy at high symmetry points in the Brillouin zone such that the energy difference matches with the peaks in $\epsilon_2(\omega)$. It is important to note that, in general, not all such transitions may be allowed due to selection rules. However, in the present case, as the states across the Fermi energy consist dominantly of ${\rm O}-2p$ and ${\rm Sb}-d$ states, the selection rules are naturally satisfied.
 The possible optical transitions in the energy range $0$ - $9$ eV have been identified and explicitly shown in the corresponding bandstructure plots. The low energy peaks correspond mainly to transitions from O-$2p$ non-bonding states close to Fermi energy in the VB to Sb-$s$ and Sc-$d$ states in the CB. On the other hand, the higher energy peaks correspond to transitions from the semi-core electrons in the VB to states in the CB. 

 With increasing distortion of the ${\rm BO_6}$ and ${\rm B'O_6}$ octahedra, the position of the lowest frequency peak $A$ varies according to the bandgap values and also becomes relatively well-resolved. The relative intensity of the $B$ and $C$ peaks also changes as their ratio decreases with decreasing temperature. For the CSS, on the other hand, the peak structure is characteristically distinct with two
 well-resolved low energy peaks. These differences arise due to relative changes in the band dispersion of the O states in the valence band, as discussed earlier (see Section 3.3.2).
 
\subsubsection{Refraction Coefficients}
\begin{figure}
 \begin{center}
 \includegraphics[angle=-90,scale=0.1125]{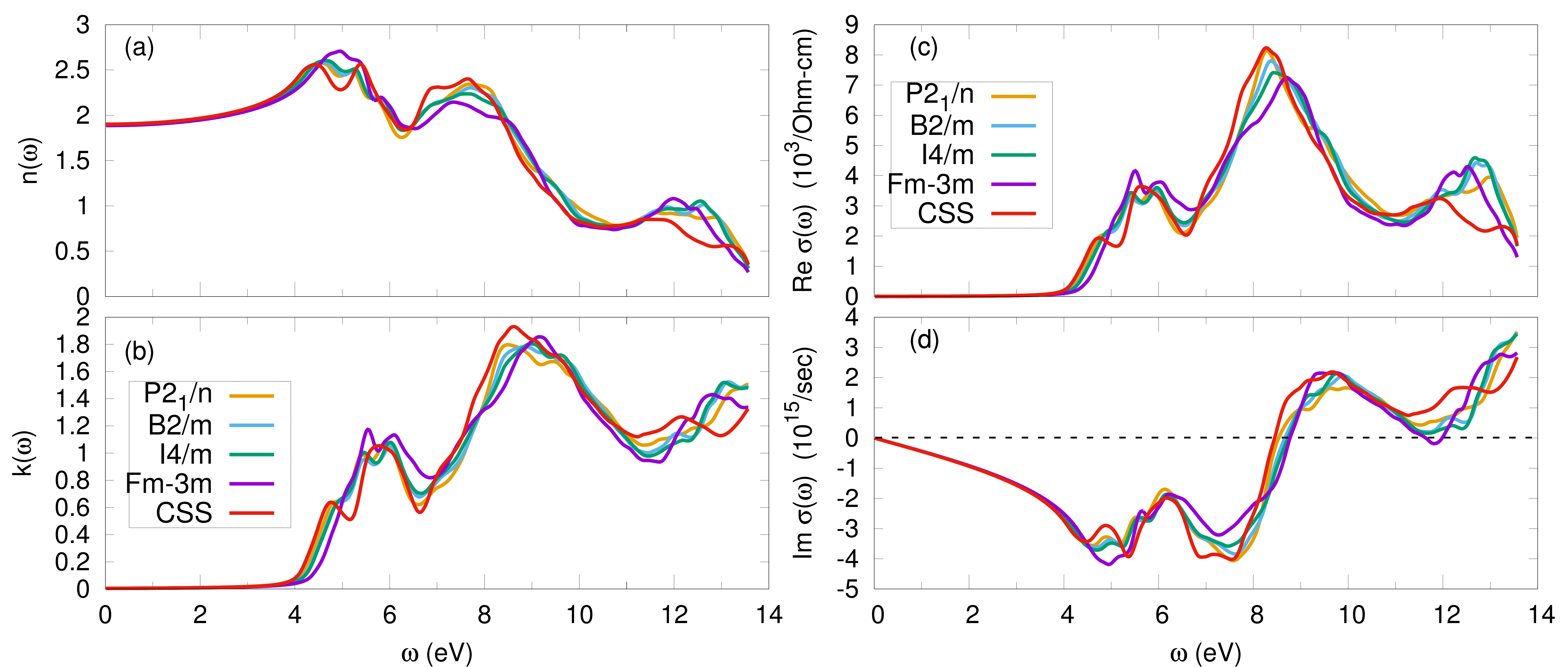}
 \end{center}
 \caption{Color Online. (a) Refractive index, (b) extinction coefficient, and the (c) real and (d) imaginary part of the optical conductivity, obtained from Eq. (\ref{eqn:nandk}) and Eq. (\ref{eqn:sigmak}), for the
 different phases of the SSS and CSS.} 
 \label{fig:refraexti}
 \end{figure}

 In the low-frequency limit ($\omega \rightarrow 0$), the magnitude of the dielectric response $\epsilon _1(0)$ corresponds to the static optical dielectric constant $\epsilon_\infty$. Using Eq. (\ref{eqn:staticdielectric}), the refraction coefficient is calculated from the dielectric response, and is found to be:
 \begin{eqnarray}
       n(0)=\sqrt{\epsilon_1(0)}&=&\sqrt{3.556}=1.886 \,\,\,\,\;{\rm
                         cubic}\,\,\,\, Fm\bar{3}m \nonumber \\
             &=&\sqrt{3.549}=1.884 \,\,\,\,\;{\rm
                               tetragonal}  \,\,\, I4/m \nonumber \\
             &=&\sqrt{3.628}=1.905 \,\,\,\,\; {\rm
                               monoclinic} \,\,\, B2/m \nonumber \\
             &=&\sqrt{3.641}=1.908 \,\,\,\,\; {\rm
                               monoclinic}\,\,\, P2_1/n  \nonumber \\
             &=&\sqrt{3.653}=1.911 \,\,\,\,\;{\rm
                         monoclinic}\,\, P2_1/n,\,\,\,\, {\rm CSS} 
       \label{eqn:staticdielectricval}
 \end{eqnarray}
 in conformity with the Fig. \ref{fig:refraexti}(a), which shows the finite-frequency refractive index and closely follows the real part of dielectric function.  Fig. \ref{fig:refraexti}(b) shows the extinction coefficient for the different phases of the SSS and CSS compound, as obtained from Eq. (\ref{eqn:nandk}), and has the same features as the imaginary part of the dielectric function. The finite-frequency refraction coefficient has maximum values of approximately 2.7 for the cubic phase, at the photon energy of approximately 5 eV. The corresponding maximum value for the distorted phases is slightly lower and follows the trend in the first peak in the real part of the dielectric constant.

 It is worth pointing out that the dielectric constant for SSS on YBCO and MgO is found to be 8.8 (with loss tangent equal to 2.1E-03) \cite{TauberAntimonyInterest5}, which is considerably larger than the values found in this work. We believe that the discrepancy is either due to disorder or influenced by the substrate.

 While the value of $n(0)_{xx}$ and $n(0)_{zz}$ for the ideal cubic phase is same, the tetragonal and monoclinic phases show deviation. Ratio of the components along different directions, $n(0)_{xx}/n(0)_{zz}$, is a measure of the optical anisotropy of the material. For the tetragonal and monoclinic phases of the SSS, this ratio is equal to $0.986$, $1.0097$ and $1.0074$ for $I4/m$, $B2/m$ and
 $P2_1/n$ phases, respectively. Therefore, as we move away from the ideal cubic symmetry, the optical anisotropy first decreases for the tetragonal symmetry and then increases for the monoclinic symmetry.
 For the CSS, the corresponding anisotropy ratio is equal to 1.0104, which is due to larger distortion in the octahedra induced by relatively smaller Ca cation. 
 
\subsubsection{Optical Transport}
 Fig. \ref{fig:refraexti}(c) $\&$ (d), respectively, shows the real and imaginary parts of the optical conductivity for different phases of the SSS and CSS, as obtained from Eq. (\ref{eqn:sigmak}).
 The low energy peaks exhibit characteristics similar to the $\epsilon_2(\omega)$. The position of the
 most prominent peak at approximately $8.6$ eV decreases monotonically with decreasing temperature or increasing distortions in the unit cell. Imaginary part of the optical conductivity is due to the dispersive part of the dielectric response. From small values of energy (frequency), it is less than zero and crosses zero around $8.6$ eV consistent with the zero-crossing of corresponding $\epsilon_1(\omega)$ curves for the respective phases.
 
\begin{figure}
 \begin{center}
 \includegraphics[angle=-90,scale=0.1406]{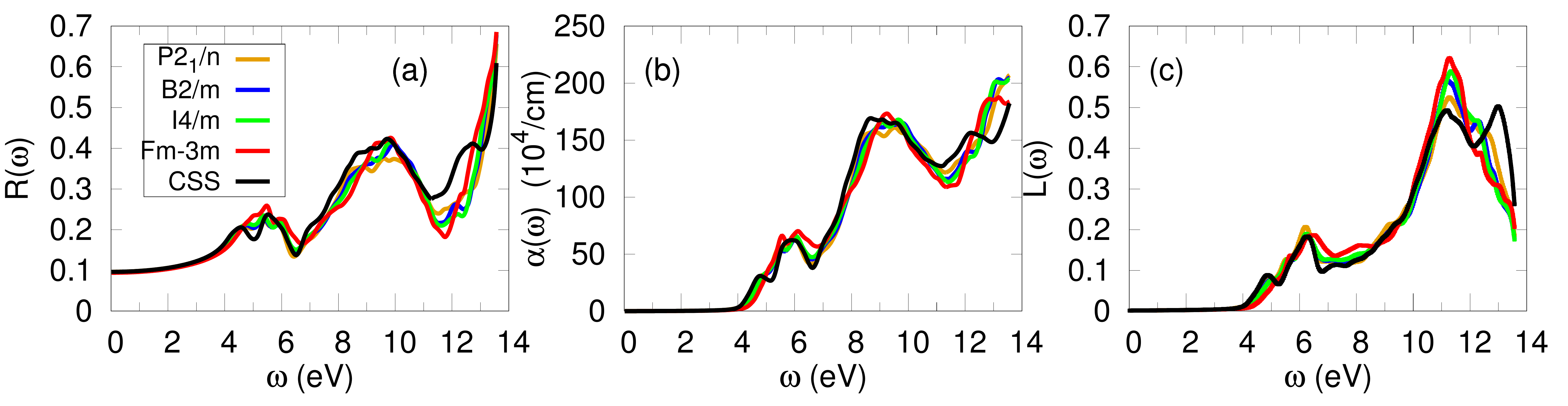}
 \end{center}
      \caption{Color Online. The finite-frequency reflectivity $R(\omega)$, absorption coefficient
      $\alpha(\omega)$ and the electron energy loss spectrum $L(\omega)$
      for the different phases of the SSS and CSS.}
      \label{fig:optiparam}
 \end{figure}

 Fig. \ref{fig:optiparam}(a) $\&$ (b) shows the frequency dependence of the reflectivity and absorption coefficient. For the absorption coefficient $\alpha(\omega)$, the absorption edge starts from
 approximately $4.0$ eV. The value of the smallest frequency at which the absorption coefficient is finite decreases monotonically from cubic $Fm\bar{3}m$ to monoclinic $P2_1/n$ phases. The peaks around the energies $4.5$, $6.0$ and $9.0$ eV arise due to inter-band transitions. The reflectance and absorption coefficient plots also suggest that these materials cannot exhibit any transparency.

\subsubsection{Electron Energy Loss Spectroscopy}
 Electron Energy Loss Spectroscopy (EELS) is considered a valuable tool for the investigation of various aspects of materials as it covers the complete energy range and includes both non-scattered and elastically scattered electrons. It is related to the energy loss as electrons traverse in the material and is usually large at the plasma frequency. At intermediate energies (typically $1$ to $50$ eV), the energy losses are
 primarily due to a mixture of single electron excitations and collective excitations (plasmons). 
 Fig. \ref{fig:optiparam}(c) shows the electron loss spectrum for the different phases of the SSS and CSS. 
 The first and second peaks, at approximately $4.5$ and $6$ eV respectively, originate from inter-band transitions from the O-$2p$ to Sb-$s$ and Sc$-d$ states, as discussed earlier. 
 
\subsection{Analysis of Symmetry-breaking Modes}

 We would now like to turn our attention to the possibility of ferroelectricity in these compounds. The key idea is that the structural distortion which relates the prototype cubic structure and the observed distorted one can be identified as the symmetry-breaking distortion. Different symmetry modes (phonon modes) responsible for multiple phase transitions in SSS have already been studied and assigned by the authors of Ref. \cite{Faik2012}. Generally, such distortions involve polar, non-polar and displacive type, leading to rich electromagnetic response in perovskites. However, in the present case, the responsible symmetry-breaking distortions are only displacive type: Let us start with the most distorted monoclinic $P2_1/n$ structure. There are two symmetry modes responsible for the transition from the (ideal) cubic $Fm{\bar 3}m$ to monoclinic $P2_1/n$, ${\rm \Gamma}^{4+}$ and ${\rm X}^{3+}$, in ($a,a,0$) and ($0,a,0$) direction, respectively. Both these modes are non-polar in nature and can be understood in terms of tilting of the octahedra \cite{Faik2012}. The mode ${\rm \Gamma}^{4+}$ corresponds to an out-of-phase rotation of the octahedra
 about the ${\rm [110]}$ direction while ${\rm X}^{3+}$ corresponds to the in-phase rotation of the octahedra about the ${\rm [010]}$ direction. On the other hand, the mode responsible for transition to tetragonal $I4/m$ and monoclinic $B2/m$ is ${\rm \Gamma}^{4+}$, in ($0,0,a$) and ($a,a,0$) direction, respectively. An illustrative visualization of these modes can be found in Ref \cite{Faik2012}. Similarly, for CSS as well, the symmetry modes in the room temperature phase are also
 non-polar, rendering both these compounds non-ferroelectric.

\section{Conclusion}
 In conclusion, we have investigated the effects of octahedral distortions on the electronic properties of the $d^0$ double perovskites SSS and CSS. We have highlighted that SSS is ideal for such a study as it undergoes multiple structural phase transitions. Both the compounds are found to be insulating in all the phases with a direct bandgap. We have also determined the bandgap of the room-temperature phases of
 these compounds experimentally, using the diffuse reflectance spectroscopy and the Kubelka-Munk spectrum.  The bandgaps for the room temperature phases of SSS and CSS is found to approximately 3.57 and 3.68 eV, respectively.

 Based on a detailed analysis of DFT calculations of the electronic and optical properties of different phases of these compounds, it was found that the isolated effect of octahedral distortions is minimal, particularly when there is large difference in the electronegativity values of the ${\rm B}$-site atoms. 
 The maximum change in the calculated value of the electronic bandgap is approximately $0.1$ eV for SSS. While for other $d^0$ perovskites, such a change in bandgap is usually accompanied by change in conduction
 bandwidth, in the present case, the conduction bandwidth remains unchanged due to large difference in the electronegativity values of ${\rm Sc}$ and ${\rm Sb}$ atoms. The electronic bandstructure plots reveal 
 that the orbitals participating in the formation of effective $t_{\rm 2g}$ band is consistent with the structural distortions of the unit cell. Changing the cation at A-site to a smaller ion increases the
 electronic bandgap by approximately $0.11$ eV consistent with the experiments as well as other ${\rm
 Sc}$ based $d^0$ double perovskites.

 The electronic bandgap values obtained from DFT calculations are in good agreement with the experimental values for room temperature phases as well as earlier studies on other Sc-based $d^0$ double perovskites.
 For SSS, although small, the bandgap initially decreases and then increases due to increase in octahedral distortions from the ideal cubic symmetry. Similar behaviour is observed for the optical anisotropy
 while the refractive index increases monotonically. These feature have been summarized in Table \ref{table:summary} for brevity. The value of dielectric response and refractive coefficient is found to
 be significantly lower than earlier reported values. Also,the non-polar nature of the primary symmetry distortion modes in these compounds suggests that these compounds are not ferroelectric.

 \begin{table}
 \centering
 \begin{tabular}{l| c| c| c| c| c|} \hline
       \multicolumn{1}{|l|}{\multirow{2}{*} {Properties}} &
       \multicolumn{4}{c|}{SSS} &
       \multicolumn{1}{c|}{CSS} \\ \cline{2-6}
       \multicolumn{1} {|c|} {} &
       \multicolumn{1} {c|}{$Fm \bar3m$} &
       \multicolumn{1} {c|}{$I4/m$} &
       \multicolumn{1} {c|}{$B2/m$} &
       \multicolumn{1} {c|}{$P2_1/n$} &
       \multicolumn{1} {c|}{$P2_1/n$} \\\cline{1-6}
 
       \multicolumn{1} {|l|} {Bandgap (eV)} &
       \multicolumn{1} {c|}{3.51} &
       \multicolumn{1} {c|}{3.41} &
       \multicolumn{1} {c|}{3.40} &
       \multicolumn{1} {c|}{3.47 (3.57)} &
       \multicolumn{1} {c|}{3.61 (3.68)} \\ 
 
       \multicolumn{1} {|l|} {Refractive Index} &
       \multicolumn{1} {c|}{1.886} &
       \multicolumn{1} {c|}{1.884} &
       \multicolumn{1} {c|}{1.905} &
       \multicolumn{1} {c|}{1.908 } &
       \multicolumn{1} {c|}{1.911} \\ 
 
       \multicolumn{1} {|l|} {Optical Anisotropy} &
       \multicolumn{1} {c|}{1.00} &
       \multicolumn{1} {c|}{0.986} &
       \multicolumn{1} {c|}{1.0097} &
       \multicolumn{1} {c|}{1.0074} &
       \multicolumn{1} {c|}{1.0104} \\ \cline{1-6}
 \end{tabular}
      \caption{Summary of the electronic and optical properties of different
      phases of the SSS and CSS. The experimental band-gap values for the room temperature phases obtained in this work are
      mentioned in brackets.}
      \label{table:summary}
 \end{table}

\section*{Acknowledgements}
 RR and MR acknowledge the funding by the European Union (ERDF) and the Free
 State of Saxony via the project 100231947 (Young Investigators Group
 Computer Simulations for Materials Design - CoSiMa) during the
 preparation of the manuscript.

\
\noindent
\textit{\bf Supplementary Material for ``Effects of octahedral tilting on the electronic structure and optical
properties of $d^0$ double perovskites $\mathbf{\rm
A_2ScSbO_6}$ ($\mathbf{\rm A=Sr, Ca}$)"}

\begin{centering}
\section*{Abstract}
\end{centering}
\noindent
In this supplementary material, we provide the optimal structural parameters used for the DFT calculations of different phases of ${\rm Sr_2ScSbO_6}$ (SSS) and ${\rm Sr_2ScSbO_6}$ (CSS) discussed in the main text. We also present a comparison of the bond lengths and bond angles with the corresponding experimental values for the room temperature phases.
\vspace{3cm}

\newpage
We begin with the optimal values of the lattice constants and corresponding atomic positions for all the structures of ${\rm Sr_2ScSbO_6}$ (SSS) and ${\rm Ca_2ScSbO_6}$ (CSS) discussed in the main text. Table \ref{table:atpos1} $\&$ \ref{table:atpos2} presents these values for the high and room temperature phases, respectively. 

 \begin{table}[ht!]
 \centering
 \begin{tabular}{|l| c| c| c|}
 \hline 
 Param. & $Fm\bar{3}m$  & $I4/m$ & $B2/m$ \\ [1.0ex]
 \hline 
 & & \\ [-1.0ex]
 a,b,c (${\rm \AA}$)& 8.1752 (=b,c) & 5.7493, 5.7493, 8.1439 & 9.9744, 8.1350, 5.7609 \\ 
 $\alpha,\beta,\gamma$ (deg) & 90, 90, 90 & 90, 90, 90 & 90, 90, 144.61 \\ 
 Sr/Ca & (0.25, 0.25, 0.25) & (0, 0.5, 0.25) & (0.5007, 0.7528, 0) \\ 
 Sb & (0.5, 0.5, 0.5) & (0, 0, 0.5) & (0, 0.5, 0) \\ 
 Sc & (0, 0, 0) & (0, 0, 0) & (0, 0, 0) \\ 
 O1 & (0.2554, 0, 0) & (0, 0, 0.2545) & (0.0484, 0.2923, 0)  \\ 
 O2 & - & (0.3019, 0.2104, 0) & (0.7571, 0.7824, 0.2441) \\ 
 \hline
 \end{tabular}
      \caption{The relaxed atomic positions of different atoms for
      higher temperature phases of the SSS compound.}
      \label{table:atpos1}
 \end{table}

 \begin{table}[ht!]
      \centering
      \begin{tabular}{|l| c| c| }
      \hline 
      Param. & SSS & CSS \\ [1.0ex]
      \hline 
      a,b,c & 5.7605, 5.7466, 8.1217 &  5.5659, 5.6808, 7.9415 \\ 
      $\alpha,\beta,\gamma$ & 90, 90.03, 90 & 90, 90.02, 90 \\ 
      Sr/Ca & (0.0049, 0.0201, 0.2499) & (0.0129, 0.0499, 0.2503)  \\ 
      Sc & (0.5, 0, 0) & (0.5, 0, 0) \\ 
      Sb & (0, 0.5, 0) & (0, 0.5, 0) \\ 
      O1 & (0.2693, 0.2804, 0.0336) & (0.2922, 0.3021, 0.0512) \\ 
      O2 & (0.2820, 0.2706, 0.4662) & (0.3059, 0.2913, 0.4464) \\ 
      O3 & (0.9361, 0.4895, 0.2442) & (0.9003, 0.4648, 0.2434) \\ 
      \hline
      \end{tabular}
            \caption{The relaxed atomic positions of the different atoms for
            the room temperature phases of the SSS and CSS compounds. At room temperature, both the compounds crystallize in ${\rm P2_1/n}$ symmetry. }
            \label{table:atpos2}
 \end{table}

\newpage
Table \ref{table:comparison} lists the bond lengths and bond angles derived from the optimal DFT structural parameters for all the phases of these compounds. A comparison of the same with the corresponding experimental values \cite{Faik2012} for the room temperature phases is also provided.

 \begin{table}[ht!]
      \centering
      \begin{tabular}{l| c| c| c| c| c|} \hline
      \multicolumn{1}{|c|}{\multirow{1}{*} {Bond
      lengths (${\rm \AA}$),}} &
      \multicolumn{4}{c|}{SSS} &
      \multicolumn{1}{c|}{CSS} \\ \cline{2-6}
      \multicolumn{1} {|c|} {Angles (deg)} &
      \multicolumn{1} {c|}{$P2_1/n$} &
      \multicolumn{1} {c|}{$B2/m$} &
      \multicolumn{1} {c|}{$I4/m$} &
      \multicolumn{1} {c|}{$Fm \bar3m$} &
      \multicolumn{1} {c|}{$P2_1/n$} \\ \cline{1-6}

      \multicolumn{1} {|c|} {Sb-O1} &
      \multicolumn{1} {c|}{2.053 (2.001)} &
      \multicolumn{1} {c|}{2.011} &
      \multicolumn{1} {c|}{2.017} &
      \multicolumn{1} {c|}{2.002} &
      \multicolumn{1} {c|}{2.021 (1.998)} \\ 

      \multicolumn{1} {|c|} {Sb-O2} &
      \multicolumn{1} {c|}{2.010 (1.975)} &
      \multicolumn{1} {c|}{2.016} &
      \multicolumn{1} {c|}{2.000} &
      \multicolumn{1} {c|}{-} &
      \multicolumn{1} {c|}{2.014 (1.993)} \\ 

      \multicolumn{1} {|c|} {Sb-O3} &
      \multicolumn{1} {c|}{2.011 (1.992)} &
      \multicolumn{1} {c|}{-} &
      \multicolumn{1} {c|}{-} &
      \multicolumn{1} {c|}{-} &
      \multicolumn{1} {c|}{2.018 (1.996)} \\ 

      \multicolumn{1} {|c|} {Sc-O1} &
      \multicolumn{1} {c|}{2.065 (2.052)} &
      \multicolumn{1} {c|}{2.097} &
      \multicolumn{1} {c|}{2.116} &
      \multicolumn{1} {c|}{2.086} &
      \multicolumn{1} {c|}{2.111 (2.092)} \\ 

      \multicolumn{1} {|c|} {Sc-O2} &
      \multicolumn{1} {c|}{2.119 (2.072)} &
      \multicolumn{1} {c|}{2.113} &
      \multicolumn{1} {c|}{2.073} &
      \multicolumn{1} {c|}{-} &
      \multicolumn{1} {c|}{2.121 (2.093)} \\

      \multicolumn{1} {|c|} {Sc-O3} &
      \multicolumn{1} {c|}{2.118 (2.062)} &
      \multicolumn{1} {c|}{-} &
      \multicolumn{1} {c|}{-} &
      \multicolumn{1} {c|}{-} &
      \multicolumn{1} {c|}{2.109 (2.079)} \\ 

      \multicolumn{1} {|c|} {$\langle$ Sc-O1-Sb $\rangle$} &
      \multicolumn{1} {c|}{160.4 (165.6)} &
      \multicolumn{1} {c|}{163.7} &
      \multicolumn{1} {c|}{159.2} &
      \multicolumn{1} {c|}{180} &
      \multicolumn{1} {c|}{146.9 (148.6)} \\

      \multicolumn{1} {|c|} {$\langle$ Sc-O2-Sb $\rangle$} &
      \multicolumn{1} {c|}{162.3 (166.7)} &
      \multicolumn{1} {c|}{158.4} &
      \multicolumn{1} {c|}{180} &
      \multicolumn{1} {c|}{-} &
      \multicolumn{1} {c|}{148.9 (149.0)} \\ 

      \multicolumn{1} {|c|} {$\langle$ Sc-O3-Sb $\rangle$} &
      \multicolumn{1} {c|}{159.1 (163.5)} &
      \multicolumn{1} {c|}{-} &
      \multicolumn{1} {c|}{-} &
      \multicolumn{1} {c|}{-} &
      \multicolumn{1} {c|}{149.1 (149.5)} \\ \cline{1-6}

 \end{tabular}
      \caption{Bond lengths and angles for different phases of SSS and CSS as
      obtained from the relaxed crystal structures. The corresponding
      experimental values are shown in brackets, as obtained from Ref.
      \cite{Faik2012}.}
      \label{table:comparison}
 \end{table}


\end{document}